\documentclass[a4paper,pre,reqno,superscriptaddress,showpacs,twocolumn]{revtex4-1}

\usepackage{graphicx,color}
\usepackage{dcolumn}
\usepackage{epsfig}

\usepackage[centertags]{amsmath}
\usepackage{amsfonts,amsmath}
\usepackage{euscript}
\usepackage{amssymb}
\usepackage{amsthm}
\usepackage{newlfont}
\usepackage{mathrsfs}
\usepackage{subfigure}

\newcommand{\opunit}{\text{1}\kern-0.22em\text{l}}

\newcommand{\ie}{\textit{i.e.}}

\newcommand{\id}{\textrm{d}}

\def\bea{\begin{eqnarray}}
\def\eea{\end{eqnarray}}
\def\ba{\begin{array}}
\def\ea{\end{array}}
\def\n{\nonumber}

\def\la{\langle}
\def\ra{\rangle}

\def\ve{\varepsilon}

\begin{document}
 
\title{Long time position distribution of an  active Brownian particle in two dimensions}
\author{Urna Basu}
\affiliation{Raman Research Institute, Bengaluru 560080, India}
\author{Satya N. Majumdar}
\affiliation{LPTMS, CNRS, Univ. Paris-Sud, Universit\'{e} Paris-Saclay, 91405 Orsay, France}
\author{Alberto Rosso}
\affiliation{LPTMS, CNRS, Univ. Paris-Sud, Universit\'{e} Paris-Saclay, 91405 Orsay, France}
\author{Gr\'egory Schehr}
\affiliation{LPTMS, CNRS, Univ. Paris-Sud, Universit\'{e} Paris-Saclay, 91405 Orsay, France}

\begin{abstract}
We study the late time dynamics of a single active Brownian particle in two dimensions with speed $v_0$ and
rotation diffusion constant $D_R$. We show that at late times $t\gg D_R^{-1}$, while the position
probability distribution $P(x,y,t)$ in the $x$-$y$ plane approaches a Gaussian form near its peak describing
the typical diffusive fluctuations, it has non-Gaussian tails describing atypical rare fluctuations when
$\sqrt{x^2+y^2}\sim v_0 t$. In this regime, the distribution admits a large deviation form, $P(x,y,t) \sim \exp\left[-t\, D_R\,
\Phi\left(\sqrt{x^2+y^2}/(v_0 t)\right)\right]$, where we compute the rate function $\Phi(z)$ analytically
and also numerically using an importance sampling method. We show that the rate function $\Phi(z)$, encoding
the rare fluctuations, still carries the trace
of activity even at late times. Another way of detecting activity at late times is to subject the active particle
to an external harmonic potential. In this case we show that the stationary distribution $P_\text{stat}(x,y)$ 
depends explicitly on the activity parameter $D_R^{-1}$ and undergoes a crossover, as $D_R$ increases, from a ring shape in
the strongly active limit ($D_R\to 0$) to a Gaussian shape in the strongly passive limit $(D_R\to \infty)$.

\end{abstract}

\maketitle

\section{Introduction} Recent years have seen enormous activities, both theoretical and experimental, in the study of the 
dynamics of self-propelled active particles. These self-propelled particles generate dissipative directed motion by consuming 
energy directly from the environment \cite{Romanczuk,soft, BechingerRev,Ramaswamy2017,Marchetti2017} and appear in a wide variety 
of biological and soft matter systems which include bacterial motion \cite{Berg2004, Cates2012}, cellular tissue behaviour 
\cite{tissue}, formation of fish schools \cite{Vicsek, fish} as well as granular matter \cite{gran1,gran2} and colloidal surfers 
\cite{cluster2}, amongst others. For interacting self-propelled particles novel collective behaviours have been observed 
such as flocking 
\cite{flocking1, flocking2}, clustering \cite{cluster1,cluster2,SEB_16,SEB_17}, phase separation \cite{separation1, separation2, separation3} 
and absence of well defined pressure \cite{Tailleur2015}. Interestingly, even in the absence of interactions, the spatio-temporal
dynamics of a {\it single} self-propelled particle exhibits rich and complex behaviour. 
This has led to a flurry of recent activities on the study of the 
stochastic processes describing the motion of a single self-propelled 
particle~\cite{BechingerRev,Potosky2012,Martens2012,ADP_2014,Solon2015,EG2015,Angelani15,Takatori,Angelani17,Malakar2018,DM_2018,ABP-pre,Franosch,EM2018,Adhar2019,GM2019,Malakar2019,Dauchot2019,LMS2019,SK2019,Sevilla2014,Sevilla2019,seifert,limmer,caprini}.

\begin{figure}[t]
 \centering
 \includegraphics[width=5.5 cm]{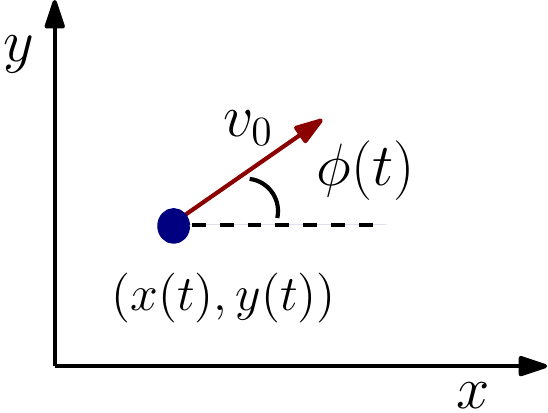}
 % curve.pdf: 0x0 px, 300dpi, 0.00x0.00 cm, bb=
 \caption{An active Brownian particle  at time $t$ moving in the $x$-$y$ plane with velocity $v_0$. The internal degree of freedom $\phi(t)$ corresponds to the orientational angle of its velocity.   }
 \label{fig:curve1}
\end{figure}

Among various models of a single self-propelled particle, perhaps one of the simplest 
is the so called active Brownian particle 
(ABP) in two dimensions. An ABP is a single overdamped particle which moves in the $2d$ $x$-$y$ plane with a constant speed 
$v_0$. In addition to its Cartesian coordinates, $(x(t),y(t))$, the particle also carries an internal ``spin'' given by the 
orientational angle $\phi(t)$ of its velocity (see Fig. \ref{fig:curve1}). This internal degree of freedom $\phi(t)$ generates 
the self-propulsion. The three coordinates $x(t)$, $y(t)$, and $\phi(t)$
  evolve with time via the coupled Langevin equations~\cite{BechingerRev,Ramaswamy2017,Marchetti2017} 
%\begin{subequations}
\bea
\dot x &=&  v_0 \cos \phi(t) \nonumber\\
\dot y &=&  v_0 \sin \phi(t)  \label{eq:phi} \\
\dot \phi &=& \sqrt{2D_R}~\eta_\phi(t)\, . \nonumber
\eea \label{eq:model}
%\end{subequations}
Here $\eta_{\phi}(t)$ is a Gaussian white noise with zero mean and a correlator $\la \eta_\phi(t) \eta_\phi(t')\ra = \delta(t-t')$. 
Thus, the orientational angle $\phi(t)$  undergoes rotational diffusion with a diffusion constant $D_R$.
In principle one can also consider an additive translational white noise with diffusion constant $D_T$ in both $x$ and $y$ equations. 
However it turns out that this additive noise does not qualitatively change the physics of the problem. Hence, for simplicity we 
drop it in the rest of the paper by setting $D_T=0$.

The angle $\phi(t)$ is just a standard one dimensional Brownian motion with auto-correlation $\la \phi(t_1) \phi (t_2) \ra = 2 D_R 
\min\{t_1,t_2\}$. Note that here the $x$ and $y$ coordinates are coupled through the angle $\phi(t)$ and hence are correlated. 
This is the origin of ``activity'' in the model. This is different from the standard ``passive'' Brownian particle (PBP) where 
the two coordinates evolve independently as $\dot x = \sqrt{2 D} \eta_x(t)$ and $\dot y = \sqrt{2 D} \eta_y(t)$, where 
$\eta_{x,y}(t)$ are independent delta correlated white noises with zero mean, and $D$ is the standard diffusion 
constant. Indeed, Eq. (\ref{eq:phi}) can be expressed in the same form as that of a PBP by writing $\dot x = 
\xi_x(t)$ and $\dot y = \xi_y(t)$, where the effective noises are $\xi_x(t) = v_0 \cos \phi(t)$ and $\xi_y(t) = v_0 \sin \phi(t).$ 
Unlike the white noises in the PBP that are independent of each other and uncorrelated in time, the active 
noises $\xi_x(t)$ and $\xi_y(t)$ are (a) correlated with each other and (b) correlated in time. For example the autocorrelation 
function of $\xi_x(t)$ is given by \cite{ABP-pre}
\begin{equation}
\label{eq:autocrr}
\langle \xi_x(t_1) \xi_x(t_2) \rangle \approx \frac{v_0^2}{2} \exp \left[ - D_R |t_1-t_2| \right]\, ,
\end{equation}
for large $t_1$ and $t_2$ with $|t_1-t_2|$ fixed and similarly for $\xi_y(t)$.  It follows from Eq.~(\ref{eq:autocrr}) that for 
times $t\gg D_R^{-1}$, the autocorrelator converges to $\langle \xi_x(t_1) \xi_x(t_2) \rangle \to 2\, D_{\text{eff}}\, \delta(t_1 -t_2)$ 
with an effective diffusion constant $D_{\text{eff}} = v_0^2/(2 D_R)$. Hence, for $t\gg D_R^{-1}$, the ABP effectively reduces to a PBP. 
Thus $D_R^{-1}$ plays the role of an `activity' parameter -- as $D_R$ increases from zero, the process crosses over 
from a strongly active regime ($D_R\to 0$) to a strongly passive one ($D_R\to \infty$).

One of the simplest and natural questions in the ABP dynamics is: how 
does the spatial distribution $P(x,y,t)$ evolve with time? 
In the case of a PBP, 
starting initially at the origin, this is simply a Gaussian at all times $t$:
\begin{equation}
P(x,y,t) = \frac{1}{4 \pi D t}\, e^{-\left(x^2+y^2\right)/4 D t}\, . 
\label{eq:diffusion}
\end{equation}
How does the presence of the internal degree of freedom $\phi(t)$ in Eq.~(\ref{eq:phi}) affect $P(x,y,t)$ ? 
In principle,  $P(x,y,t)$  can be obtained as the marginal distribution:
\bea
P(x,y,t) = \int_{-\infty}^{\infty} \id \phi~ \cal P(x,y,\phi,t)\, ,
\eea
where $ \cal P(x,y,\phi,t)$ is the probability density in the $(x,y, \phi)$ space and satisfies the Fokker-Planck equation,
\bea
\frac{\partial}{\partial t}\cal P(x,y,\phi,t) = - v_0  \bigg[\cos \phi \frac{\partial \cal P}{\partial x}+
\sin \phi \frac{\partial \cal P}{\partial y}\bigg]+ D_R \frac{\partial^2 \cal P}{\partial \phi^2}. \cr
&& \label{eq:FokkerPlanck}
\eea
However, this Fokker-Planck equation turns out to be very hard to solve explicitly. 
Thus, despite the simplicity of the ABP dynamics, 
extracting the explicit form of $P(x,y,t)$ in real space remains a challenging problem.

Recently, Kurzthaler {\it et. al.} derived \cite{Franosch} an exact expression for the Fourier transform $\la e^{-i \vec k 
\cdot \vec r(t)}\ra$, where $\vec r(t)=(x(t),y(t))$, in terms of the eigenvalues and eigenfunctions of the
Mathieu equation (see Sec. \ref{sec:model} below for details). In their derivation, 
$\langle \ldots \rangle$ includes an averaging over all possible initial 
orientations $\phi(0)$, chosen uniformly at random. Consequently, this Fourier transform 
depends only on the magnitude $k$ of the wave vector $\vec k$. However, this 
expression, although exact at all 
times, is still rather formal and inverting this Fourier transform to extract and plot $P(x,y,t)$ in the real $x$-$y$ plane is 
far from obvious.

In a recent paper \cite{ABP-pre}, using a backward Feynman-Kac approach we were able to derive, {\it for any fixed initial 
orientation} $\phi(0)$, exact and explicit expressions for the marginal distributions $P(x,t),$ $P(y,t)$ and $P(r^2,t)$ at 
short-times $t \ll D_R^{-1}.$ A fixed initial condition makes the $x$ and $y$ motion for the ABP {\it anisotropic}, especially at 
early times $t \ll D_R^{-1}$. This is manifest in the marginal distributions $P(x,t)$ and $P(y,t)$ which are completely different 
from each other at early times~\cite{ABP-pre}. For example, for the initial condition $\phi(0)=0$, it was shown that, for $t \ll 
D_R^{-1}$, the marginal distribution $P(y,t)$ has a simple Gaussian form
\begin{equation}
\label{e:Py}
P(y,t)=\frac{1}{\sqrt{2 \pi \sigma^2_y } } e^{-y^2/(2 \sigma^2_y )}, \;  \text{with} \;  \sigma_y^2 = \frac{2 v_0^2 D_R}{3} t^3 \, .
\end{equation}
In contrast, the marginal $P(x,t)$, for  $t \ll D_R^{-1}$, has a completely different expression given by the
scaling form
\begin{equation}
\label{e:Px}
P(x,t)=\frac{1}{v_0 D_R t^2 } f_x\left(\frac{v_0 t - x}{v_0 D_r t^2}\right)  \, , 
\end{equation}
where the scaling function $f_x(z)$ is non-trivial and was computed explicitly in \cite{ABP-pre}.  Note that the standard
deviation of 
$x$ grows as $t^2$, while that of $y$ grows as $t^{3/2}$, leading to anomalous super diffusion for both coordinates at early 
times. Both the anisotropy and the anomalous diffusion at early times were proposed as strong signatures of `activity' of the 
ABP dynamics~\cite{ABP-pre}.

The picture, however, is quite different, at long times $t \gg D_R^{-1}$. At long times one expects that the system forgets the 
initial condition -- hence the anisotropy disappears, and moreover by the central limit theorem normal diffusion is restored with 
an effective diffusion constant $D_\text{eff} = v_0^2/(2 D_R)$ \cite{Marchetti2017,BechingerRev,ABP-pre}. This indicates that for 
$t\gg D_R^{-1}$, the {\it typical} behaviour of $P(x,y,t)$ is described by the 
Gaussian form as in Eq.~(\ref{eq:diffusion}) with an 
effective diffusion constant $D=D_\text{eff}= v_0^2/(2 D_R)$. The question remains whether it is possible to see any signature of 
the activity in this long time regime, apart from just a trivial renormalization of the diffusion constant.

The purpose of this paper is two-fold. In the first part, we show that the same backward Feynman-Kac approach that we had used
earlier in Ref.~\cite{ABP-pre} to derive the early time dynamics ($t\ll D_R^{-1}$), can be extended to  
derive explicitly $P(x,y,t)$ 
at late times $t\gg D_R^{-1}$. We show that, while the distribution near its peak is Gaussian 
describing the probability of typical fluctuations as 
expected, it has non-trivial non-Gaussian tails describing atypical rare fluctuatuions
when $r=\sqrt{x^2+y^2}\sim v_0 t$. On this scale, we show that $P(x,y,t)$ admits a large deviation form,
\begin{equation}
\label{e:LDxy}
P(x,y,t) \sim \exp{\left[- t D_R  \,\Phi \left(\frac {\sqrt{x^2+y^2}} {v_0t}\right) \right]} \, ,
\end{equation}
where the rate function $\Phi(z)$ is supported over $z \in [0,1]$. In this paper, 
we compute $\Phi(z)$ analytically. 
Computing $\Phi(z)$ from numerical simulations is also challenging 
as it requires measuring extremely small probabilities of rare fluctuations. 
In this paper we estimate $\Phi(z)$ from numerical simulations  
with extreme precision by adapting the importance sampling method and 
find a perfect agreement with our analytical result (see Fig. \ref{fig:asym2}). 
Our main conclusion is that at late times, while there is no trace of activity in the central Gaussian peak, 
the tails of the distribution still carry signatures of activity that is encoded in the rate function $\Phi(z)$. 

We note that the marginal distribution $P(x,t)$ was studied in Ref. \cite{seifert} and a similar large deviation form
as in Eq. (\ref{e:LDxy}) was found, 
\begin{eqnarray}\label{largedev_phix}
P(x,t) \sim \exp{\left[ - t \, D_R \,\Phi_x\left( \frac{x}{v_0 t}\right) \right]} \;,
\end{eqnarray}
where the rate function $\Phi_x(z)$ is supported over $z \in [-1,1]$ and is symmetric around
$z=0$. While this rate function $\Phi_x(z)$ was found to be related to the lowest eigenvalue of the 
Mathieu equation via a Legendre transform, the asymptotic behaviors of $\Phi_x(z)$ were not extracted. 
Interestingly, the same rate function $\Phi_x(z)$ was also found in the current distribution
of an interacting active particle system \cite{limmer}, within an effective mean-field description, that just renormalises
the single particle velocity $v_0$ which now depends on the density $\rho$ -- still, the asymptotic properties of
$\Phi_x(z)$ were not analysed. In this paper, we show that, for $z \in [0,1]$, the rate function $\Phi(z)$ in Eq. (\ref{e:LDxy}) describing the 
two-dimensional probability distribution, indeed coincides with $\Phi_x(z)$ in Eq. (\ref{largedev_phix}) and we
provide the asymptotic behavior of $\Phi(z)$ both as $z \to 0$ and $z \to 1$.

In the second part of the paper we consider another way to detect the signatures of activity at late times, by 
subjecting the ABP 
to an external harmonic potential of stiffness $\mu$. Here the system approaches a stationary state at long times 
$P_{\textrm{stat}}(x,y) = P(x,y,t \to \infty).$ This stationary distribution changes its character as a function of the activity 
parameter $D_R^{-1}$. In the strongly active limit $D_R\to 0$, the stationary distribution is highly non-Gaussian and has a 
ring shape of radius $v_0/\mu$. In contrast, in the weakly active regime $D_R\to \infty$, the stationary distribution has a 
Gaussian shape. We study, both numerically and analytically, this crossover in the shape of $P_{\textrm{stat}}(x,y)$ as a 
function of the activity parameter $D_R^{-1}$.

The rest of the paper is organized as follows. In Sec. \ref{sec:model} we discuss an elegant geometrical interpretation
of the ABP dynamics in terms of a random algebraic curve in two dimensions and provide a detailed summary 
of our main results. In Sec. \ref{sec:LDF} we compute the rate function $\Phi(z)$, both analytically and numerically. Sec. 
\ref{sec:trap} is devoted to the study of the ABP in a harmonic trap. Finally we conclude in Sec. \ref{conclusion}. Some details 
of the calculations are relegated to the four Appendices.

\section{The model and the summary of the results} \label{sec:model}

The ABP model has already been defined in Eq.~(\ref{eq:phi}) in the Introduction. We assume that the particle starts at the 
origin $x=y=0,$ with a given initial orientation $\phi(0)$. We are interested in calculating the distribution $P(x,y,t)$ at late 
times $t\gg D_R^{-1}$. Before summarizing our main results, it is useful to first make a historical remark that will also provide 
an elegant geometrical representation of the ABP.

%  oriented along $x,$ \ie, with angle $\phi=0.$   

\begin{figure}[t]
 \centering
 \includegraphics[width=0.8\columnwidth]{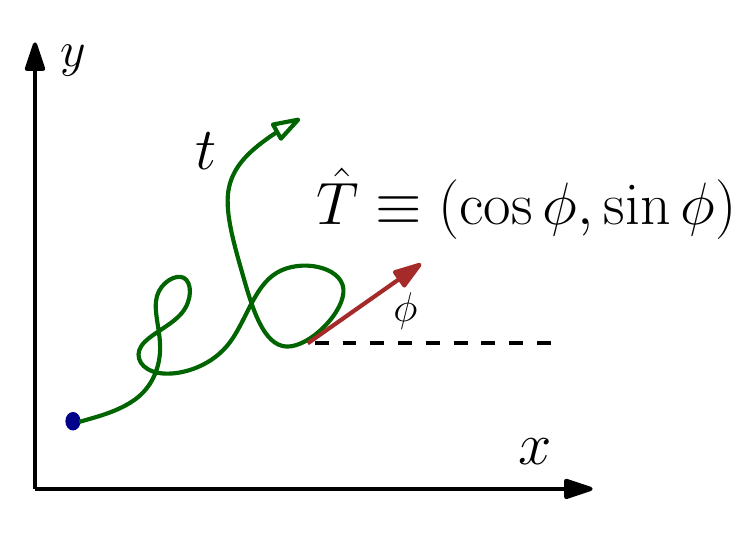}
 % curve.pdf: 0x0 px, 300dpi, 0.00x0.00 cm, bb=
 \caption{Geometric interpretation of the ABP as a two dimensional algebraic curve with random curvature \cite{Mumford}. }
 \label{fig:curve}
\end{figure}

It turns out that much before the ABP model appeared in the literature of active self-propelled particles, 
Eq.~(\ref{eq:phi}) was already introduced and 
studied in the mathematics literature by Mumford in a completely different context \cite{Mumford}. Mumford was interested in 
the properties of two dimensional random algebraic curves in the context of computer vision. Consider a continuous curve $\{ 
x(t),y(t)\}$ in $2d$, where $t$ denotes the arc length along the curve (see Fig. \ref{fig:curve} for a schematic representation).
Thus $t$ increases monotonically as one moves along the curve. 
Let $\hat T\equiv (\cos(\phi(t)),\sin(\phi(t)))$ denote the unit tangent vector to the curve at arc distance $t$, where $\phi(t)$ 
represents the angle between $\hat T$ and the $x$ axis. Hence the coordinates $(x(t),y(t))$ of the curve are expressed in terms 
of the angle $\phi(t)$ via
\bea
\label{eq:mumford}
x(t)= \int_0^t \id s \cos \phi(s)\, , \quad\quad
y(t) = \int_0^t \id s \sin \phi(s)\, .
\eea
Let $\kappa(t)$ denote the local curvature at arc distance $t$. Consequently the local radius of curvature $R(t)=1/\kappa(t)$. 
Consider an 
infinitesimal evolution of the curve from $t$ to $t+ \id t$. Clearly $R(t) \id \phi= \id t$ and this gives 
$\kappa(t)= \frac{\id 
\phi}{\id t}$. Mumford proposed a `random curvature model' for the algebraic curve 
where $\kappa(t)$ is a delta correlated white noise with zero mean. As a 
result, $\phi(t)$ in this random curvature model is just a Brownian motion with arc length $t$ playing the role of time. Hence 
the random curve described by Eq.~(\ref{eq:mumford}) is exactly equivalent to an ABP in Eq.~(\ref{eq:phi}) with $v_0=1$. Mumford 
was precisely interested in calculating $P(x,y,t)$ and wrote down the Fokker-Planck Eq.~\eqref{eq:FokkerPlanck}. However, he was 
not able to solve it and remarked ``I have looked for an explicit formula for $P$ but in vain'' \cite{Mumford}.
      
\begin{figure}[t]
 \centering
 \includegraphics[width=0.9\columnwidth]{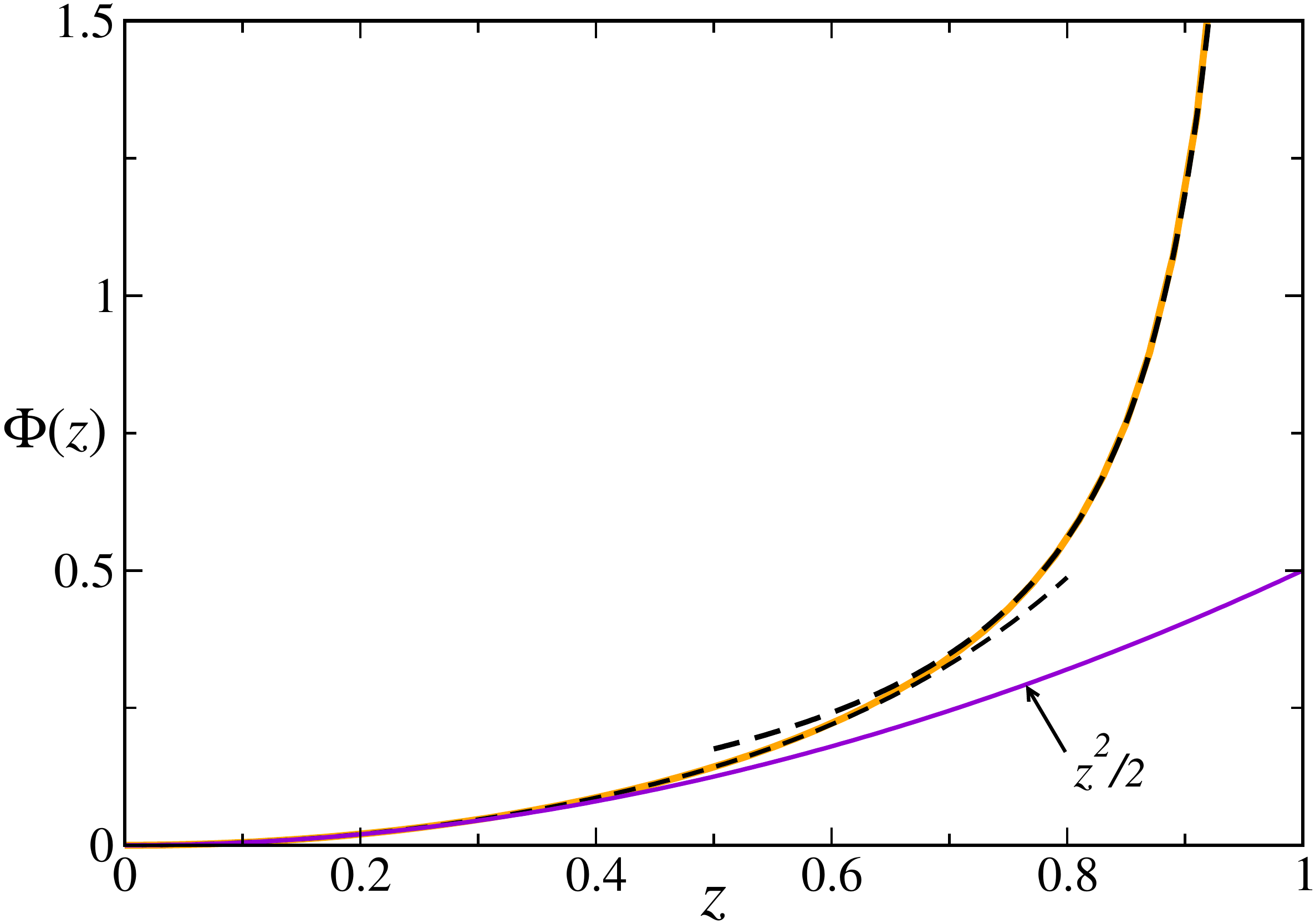}
 % curve.pdf: 0x0 px, 300dpi, 0.00x0.00 cm, bb=
 \caption{ The rate function $\Phi(z)$ supported over 
$z\in [0,1]$. The solid (orange) line corresponds to the exact 
expression given in Eq. (\ref{eq:Phiz}) evaluated using the Mathematica. The dashed lines 
correspond to the explicit asymptotic expansions in Eq. (\ref{eq:Phiz_pm1}).  
The solid (violet) line with the 
quadratic behaviour $z^2/2$ corresponds to passive Brownian 
particle.}
 \label{fig:asym2}
\end{figure}

As mentioned in the Introduction,  a recent progress was made in Ref.~\cite{Franosch}, 
where the authors derived an exact expression for the Fourier transform 
$f(k,t)=\la e^{-i \vec k\cdot \vec r(t)}\ra$, where the average is also performed over 
all initial orientations of the angle $\phi(0)$ chosen uniformly in the range $[-\pi,\pi]$.
Consequently the Fourier transform depends only on the magnitude $k$ of the wave vector $\vec k$ 
and the distribution in real space is isotropic at all times $t$.  
Ref.~\cite{Franosch} derived $f(k,t)$ in the presence of an additional translational 
noise 
in Eq. (\ref{eq:phi}). Upon setting this additional noise to zero, their result reads,
\bea
f(k,t) = \sum_{n=0}^{\infty} e^{-\lambda_{2n}t}\bigg[\int_0^{2 \pi}\frac{\id \varphi}{2\pi} 
\text{ce}_{2n}\bigg(\frac{\varphi}2, \frac{2i k v_0}{D_R}\bigg)\bigg]^2 \, ,
\label{eq:isf_franosch}
\eea
where $\text{ce}_{2n}(v,q)$ are solutions of the Mathieu equation, $\pi$-periodic and even in $v$, with eigenvalue $a_{2n}(q)$ 
(see Section \ref{sec:LDF} for more details), and $\lambda_{2n}= a_{2n}(\frac{2i k v_0}{D_R})\frac{D_R}4$. Although exact for all
$t$ in the $\vec k$ space, extracting the behavior of $P(x,y,t)$ in real space is non-trivial 
and has not been done so far.

\begin{figure*}[t]
 \includegraphics[width=0.85\linewidth]{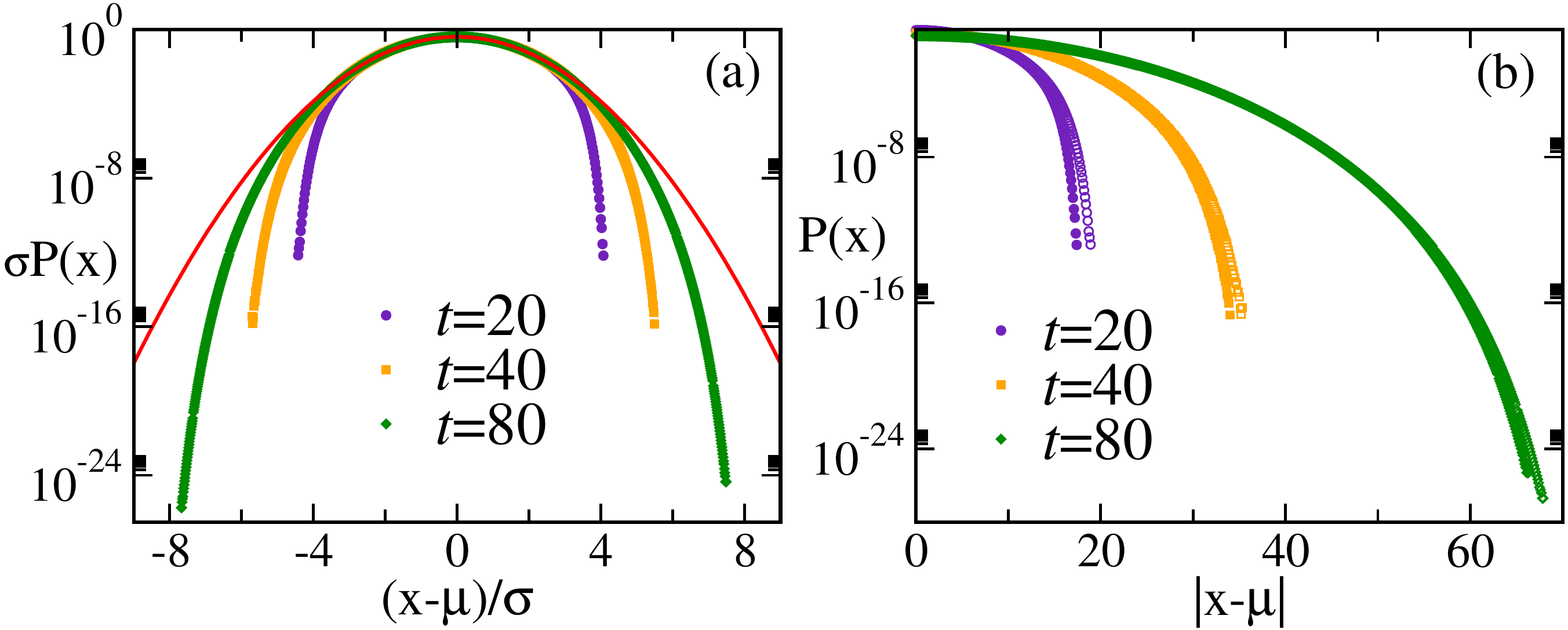}
  \caption{ Marginal distribution  $P(x,t)$ for different (large) values of $t=20,40,80$ obtained from numerical simulations 
using importance sampling. (a) The dimensionless probability density $\sigma_x P(x,t)$  
plotted as a function of the centered and rescaled position 
$(x(t) - \langle x(t) \rangle)/{\sigma_x}$, 
where, $\sigma_x=\sqrt{\langle x^2 \rangle-\langle  x\rangle^2}$ is the standard deviation. For comparison, we
have shown the pure Gaussian distribution by the solid (red) line. 
(b) $\sigma_x P(x,t)$  plotted as a function of  
$|x(t) - \langle x(t) \rangle|$ to emphasize the asymmetry between $P(x,t)$ and $P(-x,t).$ The solid and open symbols correspond to the $x(t)> \langle x(t) \rangle$ and  $x(t)< \langle x(t) \rangle$ respectively.  The two branches become identical as $t$ increases, indicating the symmetric distribution in the $t \to \infty$ limit.}
 \label{fig:sym}
\end{figure*}

In this paper, using an alternative backward Feynman-Kac formalism, 
we obtain an exact expression valid at all times for the 
moment generating function of $P(x,y,t)$---
similar to $f(k,t)$ above but with imaginary $k$. However, our solution is valid for arbitrary initial condition 
$\phi(0)$ [see Eq.~(\ref{eq:Qpt})].
Next, for large times ($t\gg D_R^{-1}$), we extract explicitly from this moment generating function the 
behaviour of $P(x,y,t)$ in real space. In particular we show that at late times the probability density 
admits a large deviation form as in Eq. (\ref{e:LDxy}) in the Introduction, that describes the
non-Gaussian tails of the distribution in the real space. We show that the
rate function $\Phi(z)$ can be expressed as the Legendre transform of the lowest eigenvalue 
$a_0$ associated with the $\pi$-periodic even solution of the Mathieu equation:
\bea
\min_{0 \le z \le 1} \left[\frac p {D_R} z + \Phi(z) \right] = \frac 14 a_0 \left(\frac{2p}{D_R}\right) \, .\label{eq:Phiz}
\eea
The rate function $\Phi(z)$  is supported over the interval $z \in [0,1]$. Note that $\Phi(z)$ depends only on 
the scaled radius $z= \sqrt{x^2+y^2}/(v_0t)$. This indicates that $-\ln P(x,y,t)$ becomes 
completely isotropic at late times and does not depend on the initial orientation $\phi(0)$. 
Hence even the marginals, such as $P(x,t),$ has the asymptotic form
\begin{equation}
\label{e:LDx2}
P(x,t) \sim \exp{\left[- t D_R  \,\Phi_x \left(\frac {x} {v_0t}\right) \right]} \, ,
\end{equation}
where $\Phi_x(z)$ is supported over the interval $z= x/(v_0 t) \in [-1,1]$ and is symmetric around $z=0$. 
For positive $z$, using the isotropy, we have  $\Phi_x(z)=\Phi(z)$, where $\Phi(z)$ is 
given in Eq. (\ref{eq:Phiz}). Thus the rate function $\Phi(z)$ can also be extracted from the late time behaviour of 
just the marginal $P(x,t)$, which turns out to be somewhat easier to analyse.
For the simplicity of notation, henceforth we will drop the subscript $x$ from $\Phi_x(z)$. As mentioned in
the introduction, this result in Eq. (\ref{e:LDx2}) already appeared in Refs. \cite{seifert,limmer}, though the detailed analytical 
form of $\Phi(z)$ was not carried out. In this paper, we show that the asymptotic behaviours of $\Phi(z)$ as $z\to 0$ and $z\to 1$ can be extracted from Eq.~(\ref{eq:Phiz}) using the known asymptotic properties of $a_0(q)$ and we obtain
\begin{widetext}
\bea
\Phi(z) = \left \{
\begin{split}
 &\frac {1}2 z^2 + \frac{7}{32} z^4 + \frac{209}{1152} z^6 +  \frac{53231}{294912} z^8 + \cdots \quad &\text{as} \quad z \to 0 \cr
& \frac 1{8(1-z)} -\frac{1}{16} - \frac{(1-z)}{64} - \frac{3}{256} (1-z)^2 - \frac{51}{4096} (1-z)^3 + \cdots \quad &\text{as} \quad z \to 1 \label{eq:Phi_boundary} \\
\end{split}
\right. \label{eq:Phiz_pm1}
\eea
\end{widetext}
A plot of $\Phi(z)$ is given in Fig.~\ref{fig:asym2}. In Sec.~\ref{sec:LDF} we further compute numerically the marginal $P(x,t)$ 
(see Fig.~\ref{fig:sym}) using an importance sampling method.  From this marginal we extract the rate function $\Phi(z)$ numerically,
as shown in Fig. \ref{fig:phiz_0}, finding excellent agreement with our analytical prediction.

In the second part of the paper we study the position distribution $P_\mu(x,y,t)$ of the ABP trapped in a harmonic potential of 
stiffness $\mu$. In this case we derive an exact recursion relation in Eq. (\ref{eq:Mkl}), valid at all times $t$, for the 
moments $M_{kl}(t) = \langle z^k(t) \bar z^l(t) \rangle$ where $z(t)= x(t)+ i y(t)$. We show that this recursion relation can be 
solved explicitly for all $t$ in the two opposite limits: (a) the strongly active limit, 
i.e. when $D_R \to 0$ and (b) the 
strongly passive limit, i.e. when $D_R \to \infty$. From these exact moments we derive, in these two limiting cases,
the exact radial distribution {\it at all times} $t$
\bea\label{Prad1}
P_{\textrm{rad}}(r,t) = \int_0^{2 \pi} P_\mu(r,\theta,t) \; d\theta \;.
\eea
where $P_\mu(r,\theta,t)$ is the position distribution in the polar coordinates.
We show that in the strongly active limit ($D_R \to 0$), the stationary position distribution at long times approaches a ring 
shape in the $x$-$y$ plane of radius $v_0/\mu$. In contrast, in the strongly passive regime ($D_R \to \infty$), the stationary 
distribution is a Gaussian
\bea\label{Pstat4}
P_{\textrm{stat}}(x,y) =  \frac{\mu D_R }{\pi v_0^2}\exp{\left[-\frac{\mu D_R (x^2+y^2) }{v_0^2} \right]} \,.
\eea
For intermediate values of $D_R$, we study the stationary distribution $P_{\textrm{stat}}(x,y)$ numerically and 
find that as $D_R$ increases, $P_{\textrm{stat}}(x,y)$ crosses over from the ring shape to the Gaussian shape 
as displayed in Fig. \ref{fig:Pxy}.

\section{Position distribution at late times: large deviation}\label{sec:LDF}

In this Section we study the behaviour of the position probability distribution $P(x,y,t)$ of an ABP at late times $t 
\gg D_R^{-1}.$ We consider Eq. (\ref{eq:phi}) and assume that the particle starts at the origin $x(0)=y(0)=0$. For a fixed 
initial orientation $\phi(0)=u$, the radial symmetry is broken and consequently the coordinates $x(t)= v_0 \int_0^t \id \tau \cos 
\phi(\tau)$ and $y(t)= v_0 \int_0^t \id \tau \sin \phi(\tau)$ will have different statistical behaviours, especially at early 
times. At late times, for typical fluctuations, this anisotropy is expected to disappear and one would recover the isotropic Gaussian 
distribution as in Eq. (\ref{eq:diffusion}) with the effective diffusion constant $D_{\text{eff}} = v_0^2/(2 D_R)$. Here we are 
interested in the atypical large fluctuations in the tails of the distribution $P(x,y,t)$, where $\sqrt{x^2+y^2}\sim v_0\, t$. 
We show that these atypical 
fluctuations also become isotropic, at least to leading order at large $t$, and $P(x,y,t)$ is described by the large deviation 
form as in Eq.~(\ref{e:LDxy}) where the rate function depends only on $z=\sqrt{x^2+y^2}/(v_0 t)$.

To compute these atypical fluctuations of $P(x,y,t)$ it turns out to be convenient to first 
study the marginal distribution $P_u(x,t)= \int \id y P(x,y,t)$, where $x(t)= v_0 \int_0^t \id \tau \cos \phi(\tau)$
and $\phi(0)=u \in (-\pi,\pi)$. For convenience, we further rescale
the $x$-cordinate and define $w(t)=x(t)/v_0= \int_0^t \id \tau\, \cos \phi(\tau)$. Therefore,
$P_u(x,t)= (1/v_0)\,  P_u(w,t)$. Thus $w(t)$ is just a functional of a one dimensional
Brownian motion $\phi(t)$ that starts at $\phi(0)=u$. The statistical properties of such Brownian 
functionals can be very conveniently
derived by using a backward Feynman-Kac approach where one treats the initial condition $\phi(0)=u$ as a variable --
for several examples and applications, see Ref.~\cite{satya_review}.
A key quantity for this method turns out to be the moment generating function
\begin{equation}
\label{e:Q}
Q_p(u,t) = \left\la e^{- p \int _0^t  d \tau  \cos \phi(\tau)} \right\ra = \int_{-t}^{t} \id w ~e^{-p w} P_u(w,t)\, , 
\end{equation}
where we note that the range of $w$ is $\in [-t,t]$. 
For the functional $w(t)= \int_0^t \id \tau\, \cos \phi(\tau)$, the backward Feynman-Kac equation 
for the moment generating function $Q_p(u,t)$ then reads \cite{satya_review},
\bea
\frac{\partial Q_p}{\partial t} = D_R \frac{\partial^2 Q_p}{\partial u^2} - p \cos u ~Q_p \, , \label{eq:Qt}
\eea
with the initial condition $Q_p(u,t=0)=1.$ Equation~\eqref{eq:Qt} is just the Schr\"{o}dinger equation in 
imaginary time for a particle in a periodic potential $\cos u.$  We look for a solution of the form $e^{-\lambda t} \psi(u).$
% \bea
% Q_p(u,t) =  e^{-\lambda t} \psi(u). \label{eq:Qsol}
% \eea
Then Eq.~\eqref{eq:Qt} becomes,
\bea\label{eq:uMathieu}
D_R \frac{\id^2 \psi}{\id u^2} + (\lambda - p \cos u) \psi(u)=0
\eea
where the physical solution should be periodic with a period $2 \pi$ (
recall that $u=\phi(0) \in (-\pi,\pi)$), 
\begin{equation}\label{eq:period}
\psi(u)=\psi(u+2 \pi) \, .
\end{equation}
It turns out that the above equation can be recast as the standard Mathieu equation \cite{Mathieu} with a rescaling $u=2 v,$ 
\bea \label{eq:Mathieu}
\psi''(v) + (a- 2 q \cos 2 v)\,\psi(v)= 0\, , 
\eea
where $a = 4 \lambda/D_R$ and $q= 2 p /D_R.$ 
Note that the periodicity condition in Eq. (\ref{eq:period}) translates in the variable $v$ to
\begin{equation}\label{eq:period2}
\psi(v)=\psi(v+ \pi) \, .
\end{equation}
For any fixed $q$, 
the Mathieu equation (\ref{eq:Mathieu}) admits four families of  
periodic solutions for a discrete set of values of the parameter $a$ called 
characteristic values, or eigenvalues \cite{Mathieu}:
\begin{itemize}
\item the elliptic cosine $\text{ce}_{2n}(v,q)$, $\pi$-periodic and even function in $v$ with eigenvalues $a=a_{2n}(q)$;
\item  the elliptic cosine $\text{ce}_{2n+1}(v,q)$, $ 2\pi$-periodic and even function in $v$  with eigenvalues $a=a_{2n+1}(q)$;
\item  the elliptic sine $\text{se}_{2n}(v,q)$, $ \pi$-periodic and odd function in $v$ with eigenvalues $a=b_{2n}(q)$; 
\item  the elliptic sine $\text{se}_{2n+1}(v,q)$, $ 2 \pi$-periodic and odd function in $v$ with eigenvalues $a=b_{2n+1}(q)$.  
\end{itemize}
 The condition in Eq. (\ref{eq:period2}) allows only the  $\pi$-periodic solutions, $\text{ce}_{2n}(v,q)$ and $\text{se}_{2n}(v,q)$.
 The general solution of Eq. (\ref{eq:uMathieu}), satisfying Eq. (\ref{eq:period}), can then be written as,
\bea
Q_p(u,t) &=& \sum_{n=0}^\infty A_{2 n}  \text{ce}_{2n} \bigg(\frac{u}{2},\frac {2p}{D_R} \bigg) e^{-\frac{t D_R}{4} a_{2n} \left(\frac {2p}{D_R}\right)}  \nonumber\\
&&+   \sum_{n=0}^\infty   B_{2 n}  \text{se}_{2n} \bigg(\frac{u}{2},\frac {2p}{D_R}\bigg) e^{-\frac{t D_R}{4} b_{2n} \left(\frac {2p}{D_R}\right)}                                     
\nonumber
%\label{eq:Qpt}
\eea
The coefficients $A_{2n}$ and $B_{2n}$ can be determined from the initial condition. Setting $t=0$  and using the orthogonality of the elliptic cosine and sine functions one gets:
\bea
 A_{2 n} \propto \int_{-\pi}^{\pi} Q_p(u,0)  \text{ce}_{2n} \bigg(\frac{u}{2},\frac {2p}{D_R}\bigg)  \id u \\
  B_{2 n} \propto \int_{-\pi}^{\pi} Q_p(u,0)  \text{se}_{2n} \bigg(\frac{u}{2},\frac {2p}{D_R}\bigg) \id u \,,
 \label{eq:coeff}
\eea
where the proportionality factors are just the normalization of $\text{ce}_{2n}$ and $\text{se}_{2n}$ respectively.
Using the initial condition $Q_p(u,t=0)=1$ and the parity of the functions $\text{ce}_{2n}$ and $\text{se}_{2n}$ , 
we immediately see that $B_{2n}=0$ while $A_{2n}\ne 0$. Therefore we finally get
  \bea
Q_p(u,t) = \sum_{n=0}^\infty A_{2 n}  \text{ce}_{2n} \left(\frac{u}{2},\frac {2p}{D_R} \right) e^{-\frac{t D_R}{4} a_{2n} \left(\frac {2p}{D_R}\right)}  \, . \label{eq:Qpt}
\eea
Clearly the dependence on the initial condition $u$ appears only in the eigenfunctions $\text{ce}_{2n}$, {\it but not in the eigenvalues} 
$ a_{2n}$.
 
At late times $t \gg D_R^{-1}$,  one can make progress since the solution in Eq. (\ref{eq:Qpt})
is dominated by the smallest eigenvalue $a_0(q)$ and we get
\bea
Q_p(u,t) \sim \exp{\left[-\frac{t D_R}4 a_0 \left(\frac {2p}{D_R}\right) \right]}. \label{eq:Q0_t}
\eea
It is important to remark that in this limit the argument of the exponential is independent of the initial condition $u$ (only 
the prefactor, which is sub-dominant in $t$, depends on $u$). The behaviour of $a_0(q)$ is known both for $q \to 0$ and $q\to \infty$ 
limits:
\bea
a_0(q) = \left \{\begin{split}
          \sum_{n=1} \alpha_{2n}\,q^{2n} &\qquad \text{for} \;\; q \to 0\cr \\[0.1 em]
          \sum_{n=0} \beta_n \, q^{1- \frac n2} &\qquad \text{for}\;\; q \to \infty \;. \label{eq:a0_qsmall}
         \end{split}
\right.
\eea
Explicit values of $\alpha_{2n}$ and $\beta_n$ are known \cite{Mathieu} and are quoted in the Appendix \ref{sec:a0}. This 
allows us to extract both the cumulants and the large deviation function of the $x$-coordinate of the particle position, as we now 
show.

Let us recall that $Q_p(u,t) =\langle \exp\left(- p \, x(t)/v_0 \right) \rangle$ is the moment generating function of $w = 
x/v_0.$ More precisely, expanding the $\ln Q_p(u,t)$ in powers of $p$ gives
\bea
\ln Q_p(u,t)= \sum_{n=1}^{\infty} \frac{(-p)^n}{n!}\, 
\frac{{\langle x^n\rangle}_c}{v_0^n}\, ,
\label{cumul.1}
\eea
where ${\langle x^n\rangle}_c$ denotes the $n$-th cumulant of $x$.
To leading order in large $t$, taking the logarithm of 
$Q_p(u,t)$ in Eq. (\ref{eq:Q0_t}) 
gives
\bea
\log Q_p(u,t) \approx - \frac{tD_R}4 a_0 \left(\frac {2p}{D_R}\right)\, . 
\label{log-Q_p}
\eea
Next we use the small $p$ expansion of $a_0\left(\frac {2p}{D_R}\right)$
in Eq. (\ref{eq:a0_qsmall}) and match powers of $p$ to extract the 
cumulants in Eq. (\ref{cumul.1}).
To leading order in large $t$, this gives for the even cumulants
\bea
\la x^{2n} \ra_c \approx - \alpha_{2n}\, \frac{(2n)!}{4}\left(\frac{2 
v_0}{D_R}\right)^{2n}\, D_R\, t \, ,
\label{even_cumul.2}
\eea
while the odd cumulants vanish to this leading order of $t$. Note that at this
leading order for large $t$, the cumulants are already independent of the
initial condition $u$.
Using the known explicit values of $\alpha_{2n}$~\cite{Mathieu}, we 
get
the first few cumulants explicitly to leading order for large $t$: 
 \bea
\la x^{2} \ra_c & \approx&  \left(\frac{ v_0}{D_R}\right)^{2} D_R\, t \nonumber \\
\la x^{4} \ra_c & \approx & -\frac{21}{4} \left(\frac{ v_0}{D_R}\right)^{4} D_R\, t  
\\
\la x^{6} \ra_c & \approx& 145 \left(\frac{ v_0}{D_R}\right)^{6} D_R\, t \nonumber \\
\la x^{8} \ra_c & \approx & -\frac{2404045}{256} \left(\frac{ v_0}{D_R}\right)^{8} 
D_R\, t \,, \nonumber 
\eea
in agreement with the results obtained in Refs. \cite{seifert,limmer} using the tilt-operator method due
to Lebowitz and Spohn \cite{LS99}, which, for Brownian motion, is equivalent to the forward Feynman-Kac formalism. Note that the sign of the even cumulants oscillate with increasing $n$. Since the odd cumulants vanish in this long-time limit, they leave no trace of the initial anisotropy for large $t$.
However, the presence of non-zero higher order even cumulants already indicates that the tails of the distribution are non-Gaussian.

To extract the large deviation behaviour of the marginal $P(x,t)$ from its moment generating function in Eq. (\ref{eq:Q0_t}) we 
proceed as follows. For fluctuations on a scale $x\sim v_0\, t$, we anticipate the large deviation form (to be verified a posteriori)
\begin{equation}
\label{PxLD}
P(x,t) \sim \exp{\left[- t\, D_R\,  \,\Phi \left(\frac x {v_0t}\right) \right]}
\end{equation}
where $\Phi(z)$ is the rate function, supported over the interval $ z \in(-1,1).$
In terms of the rescaled variable $z=x/(v_0 t)=w/t$, the moment generating function
is $Q_p(u,t)= \langle \exp\left( -p \,  x /v_0\right )\rangle= \langle \exp\left( -p \, t \, 
z\right )\rangle$.  Substituting the 
anticipated form Eq.~(\ref{PxLD}) for $P(x,t)$ (or equivalently for $P(z,t)$) in Eq.~\eqref{e:Q} gives
\begin{equation}
Q_p(u,t) \sim \int_{-1}^{1} \id z~ \exp{\bigg\{- t D_R \left[\frac p {D_R} z + \Phi(z) \right] \bigg\}} 
 \, .\label{eq:Q_phi}
\end{equation}
For large $t$, evaluating the integral by the saddle point method we get,
\begin{equation}
Q_p(u,t) \sim  \exp{\bigg\{- t\, D_R\, \min_{-1 \le z \le 1} \left[\frac p {D_R} z + \Phi(z) \right] \bigg\}} \, .\label{eq:Q_phi2}
\end{equation}
Comparing Eq. (\ref{eq:Q0_t}) and Eq. (\ref{eq:Q_phi2}), we get,
\begin{equation}
\min_{-1 \le z \le 1} \left[\frac p {D_R} z + \Phi(z) \right] = \frac 14 a_0 \left(\frac{2p}{D_R}\right).\label{eq:Phiz_min}
\end{equation}
Inverting this Legendre transform, the rate function $\Phi(z)$ can be finally expressed as
\begin{equation}
\label{eq:Phiz2}
\Phi(z) =  \max_{p} \bigg[\frac 14 a_0\left(\frac{2p}{D_R} \right) - \frac{zp}{D_R} \bigg]\, .
\end{equation}
The eigenvalue $a_0(q=2 p/D_R)$ is a symmetric function of $q$. Hence it follows immediately 
from Eq. (\ref{eq:Phiz2}) that $\Phi(z)=\Phi(-z)$. While Eq. (\ref{eq:Phiz2}) has appeared before in the
literature \cite{seifert,limmer}, its behavior for $z \to 0$ as well as $z \to \pm 1$ was not extracted. Here 
we use the asymptotic expansions of $a_0(q)$, both for small and large $q$ in Eq. (\ref{eq:a0_qsmall}), to determine
the limiting behviors of $\Phi(z)$ as $z\to 0$ and $z\to 1$ respectively. 
The details are provided in the Appendix \ref{sec:A_Phiz} and the explicit limiting behaviors of $\Phi(z)$ 
are given in Eq. (\ref{eq:Phiz_pm1}) of Sec. \ref{sec:model}. In Fig.~\ref{fig:asym2}, we provide a plot of
$\Phi(z)$ for $z\in [0,1]$ (note that $\Phi(z)=\Phi(-z)$).

So far we have studied the marginal $P(x,t)$ and observed that to leading order in large $t$, $-\ln P(x,t)$ is independent of the 
initial condition $u$. This means that one would observe the same rate function 
$\Phi(z)$ for the marginal distribution along any 
axis, and not just for $P(x,t)$. As a consequence $-\ln P(x,y,t)$ would also be described by the same rate function $\Phi (z)$
with $z=\sqrt{x^2+y^2}/(v_0 t) \in (0,1)$. This gives the large deviation form for $P(x,y,t)$ as announced in 
Eq. (\ref{e:LDxy}) 
in the Introduction. The rate function $\Phi (z)$ associated with $P(x,y,t)$ is thus
the same as in Eq. (\ref{eq:Phiz2}), but with its argument $z \in (0,1)$.
Note that the large deviation form in Eq. (\ref{e:LDxy}) not only contains the probability of extremely large fluctuations of 
order $r = \sqrt{x^2+y^2} \sim v_0 t $, but also the typical fluctuations where $r \sim \sqrt{t}$. To see this, we note that for 
$r \sim \sqrt{t}$ , the scaled variable $z = r/( v_0 t)\sim O(1/\sqrt{t})$ and hence is very small for large $t$. Using the 
quadratic form of $\Phi(z) \sim z^2/2$ near $z=0$ in the small $z$ expansion in 
Eq. (\ref{eq:Phiz_pm1}) and substituting this in Eq. 
(\ref{e:LDxy}), one recovers the typical Gaussian fluctuations
\begin{equation}
P(x,y,t) \sim e^{-\left(x^2+y^2\right)/4 D_{\text{eff}} t}. \label{eq:eff_diffusion}
\end{equation}
with $D_{\text{eff}} =v_0^2/(2 D_R)$.

We close this discussion with a final remark. We note that the result for $\Phi(z)$ in 
Eq.~(\ref{eq:Phiz2}) could also have been
derived directly from the result of Kurzthaler et. al. in Eq.~(\ref{eq:isf_franosch}). 
However, we presented here an
alternative derivation based on the backward Feynman-Kac approach for two reasons. First, our result in Eq. (\ref{eq:Qpt}) is
valid for arbitrary initial condition $\phi(0)$ and demonstrates, in particular, how the dependence on the initial condition
dispapears at late times, leading to an isotropic tail of the position distribution $P(x,y,t)$ in the $x$-$y$ plane in
Eq.~(\ref{e:LDxy}), with a rate function that only depends on the rescaled radial 
distance $z=\sqrt{x^2+y^2}/(v_0 t)$.
Secondly, we wanted to develop a single unifying method that is able to provide explicitly $P(x,y,t)$ both at early times
$t\ll D_R^{-1}$~\cite{ABP-pre} as well as at late times $t\gg D_R^{-1}$ -- our approach based on the backward Feynman-Kac formalism 
does exactly that.

\subsection{Numerical Measurement of the Large Deviation Function}

The dynamics of the ABP is also simulated numerically to measure the position probability distribution and the large deviation 
functions. For measuring the moments and distributions in the typical regime one can use the standard Euler's method where the 
Langevin equations are discretized as,
\bea
x(t+ \id t) &=& x(t) + v_0 \cos \phi(t) \id t \cr
y(t+ \id t) &=& y(t)+ v_0 \sin \phi(t) \id t \cr
\phi(t+ \id t) &=& \phi(t) + \sqrt{2 D_R \id t} ~\eta_\phi(t) 
\eea
where $\eta_\phi(t)$, for each $t$, is an independent random number drawn from the zero mean unit variance Gaussian distribution.  
We start from the fixed initial condition $\phi(0)=0$, set $\id t=10^{-3}$ and $10^{-4}$, $v_0= D_R=1$ and measure only the 
marginal distribution $P(x,t)$ for different values of $t$ (for this we do not need to monitor the $y$ coordinate).  Since 
$\phi(0)=0$, the average value of $x$ is nonzero and is given by \cite{ABP-pre}
\begin{equation}
\label{xmean}
\langle x (t) \rangle =\langle \cos \phi(t)\rangle = \frac{v_0}{D_R} \left ( 1-e^{-D_R t} \right).
\end{equation}
Note that when we monitor $P(x,t)$ we actually plot it as a function of $x(t) -\langle x(t) \rangle$.

Using this standard Euler method of integrating the Langevin equation, we can easily 
sample $10^8$ realizations. This limits the 
smallest probabilities which can be resolved to be $> 10^{-8}.$ 
To estimate 
$P(x,t)$ when its value is much smaller, e.g., when $P\sim 10^{-25},$ we use the {\it Importance Sampling} method. 
This approach has been 
successfully used to extract the tails of distributions with extremely small probabilities
in a wide variety of problems \cite{IS1,IS2,ISRMT1,ISRMT2,ISCH1,ISCH2,ISCH3,ISCH4,ISCH5,ISKPZ2,ISLIS}.
The basic idea behind the importance sampling method is to sample trajectories (or 
configurations in general) ending at 
$x(t)$ with an additional exponential tilt $e^{-\theta x(t)}$, where $\theta$ is 
an adjustable parameter. Positive values of $\theta$ will bias the trajectories with very negative $x(t) \simeq - t.$ Contrarily, 
negative $\theta$ samples trajectories ending near the other limit, \ie, $x(t) \simeq t.$

Let ${\cal P}(\omega)$ denote the probability of the trajectory $\omega = \{ x_s; 0 \le s \le t \}$ of the ABP during the time interval $[0,t].$ The expectation value of any observable $O(\omega)$ is given by 
\bea
\la O(\omega) \ra_{\cal P} = \int \cal D \omega ~ {\cal P}(\omega)\,O(\omega).
\eea
The presence of the tilt introduces a bias in the trajectory probabilities,
\bea
{\cal Q}(\omega) = {\cal P}(\omega) \frac{e^{-\theta x(t)}}{Z_\theta}
\eea
where $Z_\theta$ is the normalization constant which depends only on $\theta$ and $t.$ The expectation value $\la O(\omega) \ra_{\cal P}$ can be computed from this tilted ensemble by reweighing the observable,
\bea
\la O(\omega) \ra_{\cal P} = \int \cal D \omega~ \tilde O(\omega){\cal Q}(\omega) 
\eea
where 
\bea
\tilde O(\omega) = \frac{O(\omega){\cal P}(\omega)}{{\cal Q}(\omega)} = e^{\theta x(t)} Z_\theta O(\omega).
\eea

In practice, a trajectory is completely specified by a sequence of $N=t/dt$ Gaussian random numbers $\eta_i.$ In order to generate trajectories from the biased ensemble we rely on a Metropolis approach. Starting from an allowed trajectory $\omega$ ending at $x(t)$ we generate a trial tilted trajectory $\tilde \omega$ by modifying $r$ fraction of  the random numbers. The trial trajectory is accepted with a probability $P_\text{Met}=\min(1, e^{-\theta (\tilde x(t)-x(t))})$ where $\tilde x(t)$ denotes the ending point of the trial trajectory. The value of the parameter $r$  is adjusted in order to have $P_\text{Met} \approx 0.5$ in average.

\begin{figure}[t]
 \centering
 \includegraphics[width=0.9\columnwidth]{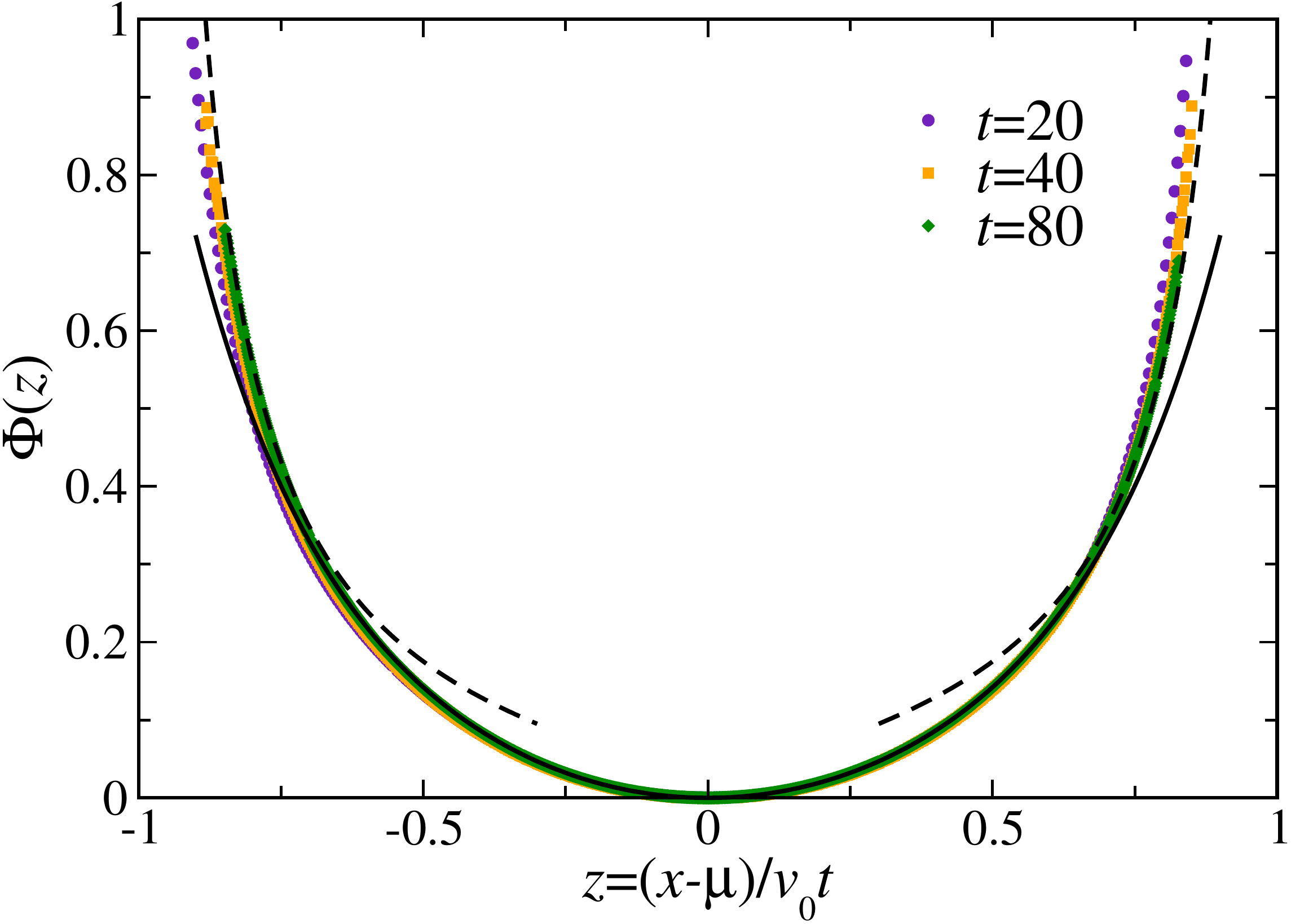}
 % AB_cumul.pdf: 638x487 pixel, 72dpi, 22.51x17.18 cm, bb=0 0 638 487
 \caption{The large deviation function $\Phi(z)$ vs $z=(x(t)-\la x(t)\ra)/t$ for three different values of $t,$ as obtained from numerical simulations. The solid black line corresponds to the asymptotic behaviour near  $z\to 0$ in  Eq.~\eqref{eq:Phiz_pm1}. The dashed black lines correspond to  asymptotic behaviour near  $z\to \pm 1$  in  Eq.~\eqref{eq:Phiz_pm1}.
 }
 \label{fig:phiz_0}
\end{figure}

To measure the distribution $P(x,t)$ for a wide range of values of $x,$ we change the value of the parameter $\theta.$ In the data presented in Figs. \ref{fig:sym} and \ref{fig:phiz_0} we have used $\theta= \pm 0.75, \pm 1.5, \pm 2.0$ and $\pm 2.5.$ 
The histogram obtained for each value of $\theta$ is shifted by an unknown amount $Z_\theta.$ To fix it, we use the histogram obtained from the standard Euler simulation, which is correctly normalized and accurate near the origin $x=0$, and corresponds to $\theta=0.$
For the smallest negative (positive) value of $\theta,$ we match the histogram obtained from 
the biased sampling with the right (left) part of the $\theta=0$ curve. We continue the same matching procedure for the subsequent values of $\theta$ to get the full curve $P(x,t).$

The marginal distribution $P(x,t)$ thus obtained for different values of $t$ are plotted in Fig.~\ref{fig:sym}(a). As is visible from this plot, the importance sampling has allowed us to resolve $P(x,t)$ near the boundaries $x = \pm 1$ to an accuracy smaller than $10^{-25}$ for $t =20.$ %Even for the curve corresponding to the largest $t=80,$ the $P(x,t)$ could be measured upto an accuracy $10^{-14}.$
Fig. \ref{fig:sym}(b) shows  $P(x,t)$ plotted as a function of $|x(t)-\langle x(t) \rangle|$ which illustrates that the distribution becomes symmetric around the mean as $t$ increases.

The large deviation function $\Phi(x/v_0 t)$ is extracted from the $P(x,t)$ obtained from numerical simulations following,
\bea
\Phi(x/v_0 t) = - \frac 1{D_R t} [\log P(x,t) - \log P(0,t)] 
\eea
This ensures that $\Phi(0)=0.$ This is plotted in Fig \ref{fig:phiz_0} for different (large) values of $t.$  
The symbols correspond to the data obtained from numerical simulations and lines correspond to the asymptotic expansions of the rate function $\Phi(z)$ in Eq. \eqref{eq:Phiz_pm1}.  
The agreement between the numerical data and the analytical curves, both near $z=0$ and $z = \pm1,$ improves as $t$  increases, validating our prediction.

The non-trivial behaviour of the large deviation function is one clear sign of `activeness' of ABP at late times. As already mentioned, another, more direct, way to explore the `active' regime is to put the ABP in an external potential. In the next section we investigate the behaviour of an ABP in a harmonic potential.

%\section{Numerical Methods}

%There are still some asymmetry visible for large values of $z,$ which decreases as $t$ is increased.

\section {ABP in a harmonic trap} \label{sec:trap}

In this section we consider the behaviour of an ABP in the presence of a confining harmonic potential $U(x,y) =\mu(x^2+y^2)/2$. 
In this case, the Langevin equations governing the dynamics of the particle become,

\bea
\dot x &=& -\mu x+ v_0 \cos \phi(t)  \nonumber\\
\dot y &=& -\mu y + v_0 \sin \phi(t)  \label{eq:trap_modely} \\
\dot \phi &=& \sqrt{2D_R}~\eta_\phi(t). \nonumber
\eea

\begin{figure}[b]
 \centering
% \vspace*{-0.5 cm}
\hspace*{-0.4cm} \includegraphics[width=2.7cm]{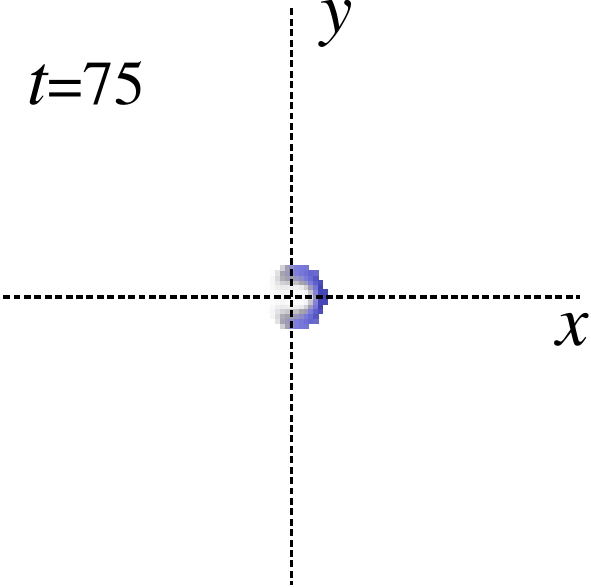}\hspace*{0.1cm}\includegraphics[width=2.7cm]{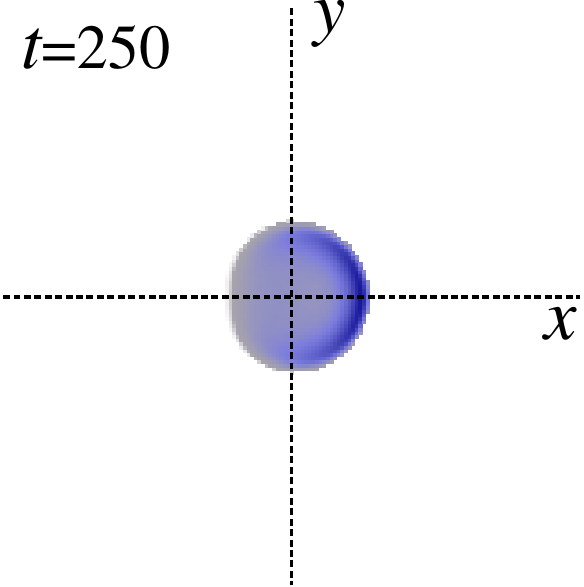}\hspace*{0.1cm}\includegraphics[width=2.7cm]{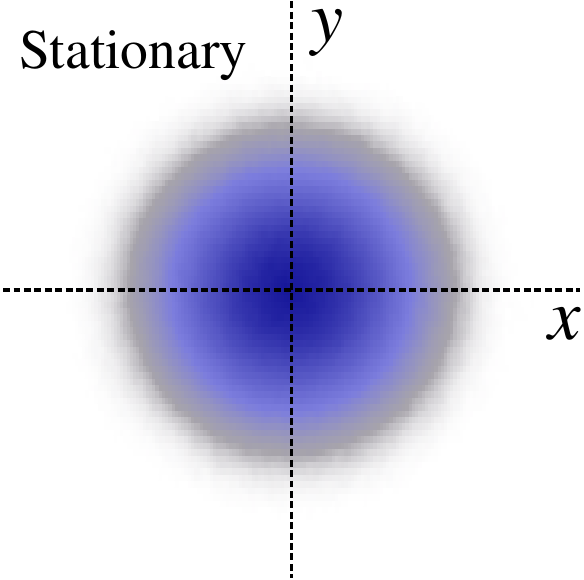}
%\vspace*{-0.5 cm}
 \includegraphics[width=8.7cm]{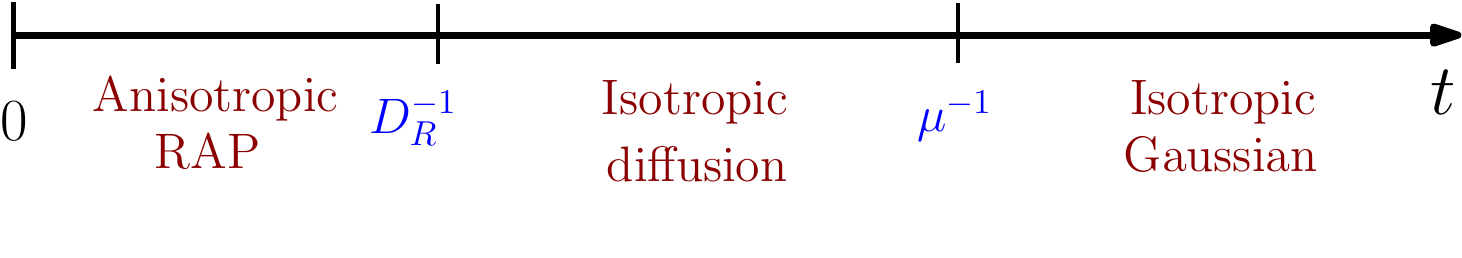}
% \vspace*{0.5 cm}
\hspace*{-0.2cm} \includegraphics[width=3.0cm]{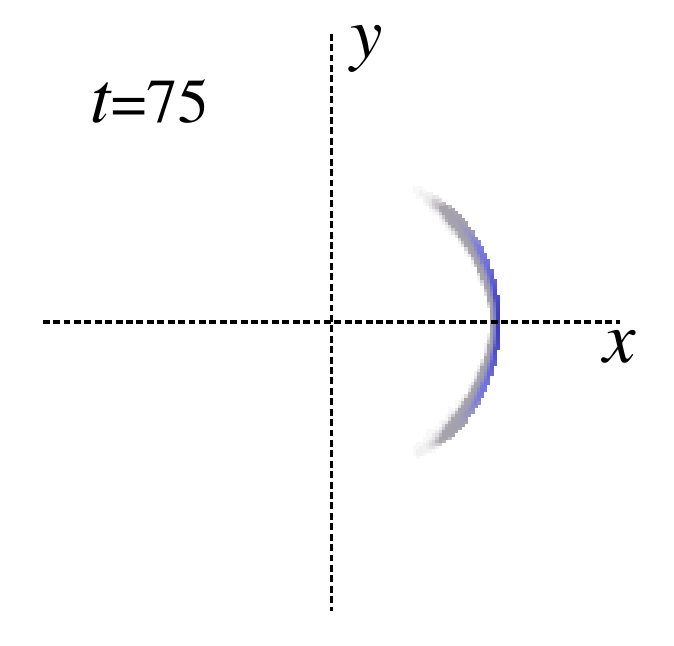}\hspace*{-0.2cm}\includegraphics[width=3.0cm]{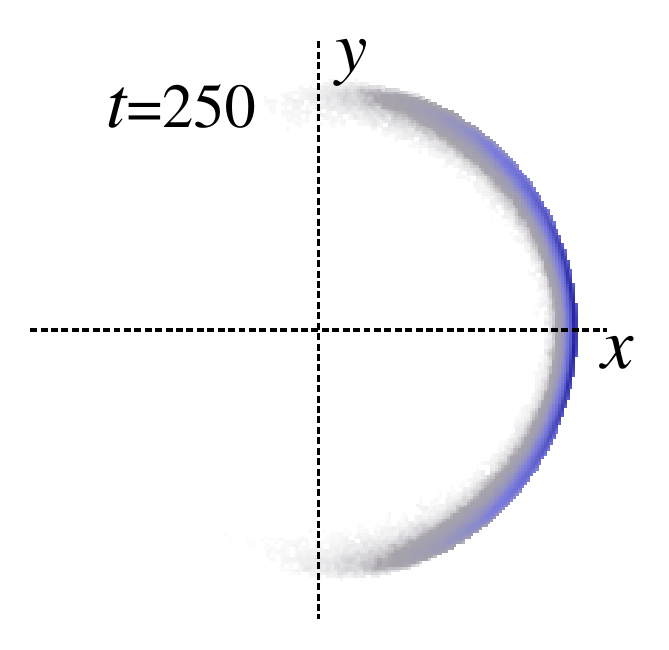}\hspace*{-0.2cm}\includegraphics[width=3.0cm]{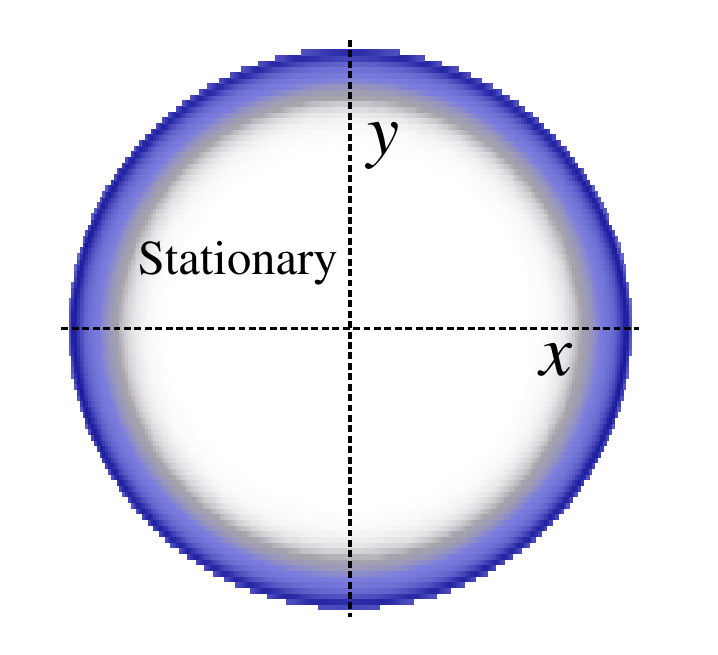}
%\vspace*{-1.5 cm}
 \includegraphics[width=8.7cm]{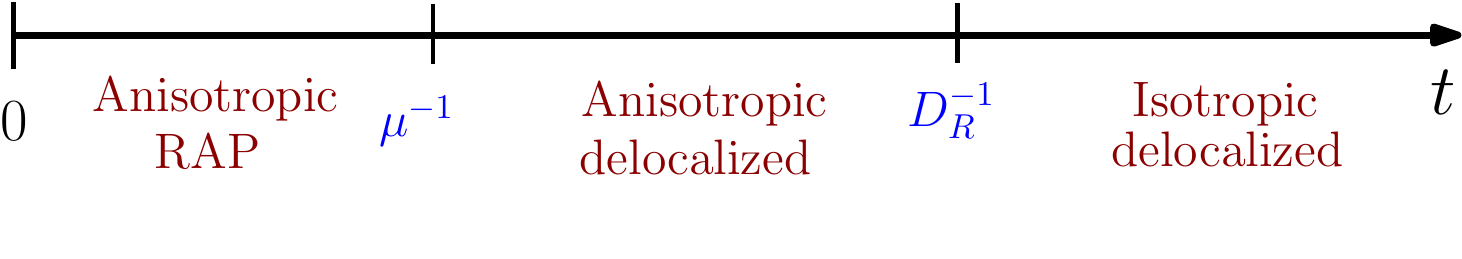}
 \caption{Position probability distribution $P(x,y,t)$ for an ABP in a  $2d$ harmonic trap of strength $\mu$ at different time $t.$  Upper and lower panels correspond to the cases $ \mu^{-1}> D_R^{-1}$ and $ \mu^{-1} < D_R^{-1},$ respectively. The presence of anisotropy at short-times and the delocalized stationary state (for $ \mu^{-1} <D_R^{-1}$) are two specific signatures of activity.
 The numerical data have been obtained for $D_R^{-1}=10^2$  and $ \mu^{-1}=10^3$ (upper panel) and $D_R^{-1}=10^3$ and $ \mu^{-1}=10^2$ (lower panel). }
 \label{fig:snap1}
\end{figure}

The ABP in a harmonic trap has been extensively studied both theoretically and experimentally ~\cite{Takatori, Solon2015, 
Potosky2012, Dauchot2019, Malakar2019,caprini}. In a recent experiment, Janus swimmers were confined in a two-dimensional harmonic-like 
trap with the use of an acoustic tweezer and the stationary density was measured by varying the trap strength~\cite{Takatori}. 
Strong signatures of activity were observed even in the dilute limit, with a crossover from a Gaussian-like stationary state, to 
a strongly active stationary state, where the particles cluster at the outskirts of the trap. The dilute limit corresponds to a 
collection of non-interacting Active Brownian Particles (ABP) in a harmonic potential as in Eq. (\ref{eq:trap_modely}).
Numerical studies of this model have also observed a similar crossover in the stationary state \cite{Solon2015,Potosky2012}.

Dynamical behaviour of an ABP differs crucially from that of a PBP, also in the presence of a harmonic potential.  For a 
`passive' or ordinary Brownian particle, the presence of a harmonic trap of strength $\mu$ sets a relaxational time scale 
$\mu^{-1}.$ At times $t \ll \mu^{-1}$, the particle diffuses isotropically and for $t \gg \mu^{-1}$, a Gaussian (Boltzmann) 
stationary distribution is reached. As explained before, for an ABP, the coupling to the rotational diffusion introduces an 
additional time scale $D_R^{-1}$, where $D_R$ is the rotational diffusion constant.

While the activity induced crossover in the stationary position distribution of an ABP has been studied both experimentally and 
numerically, the interplay of the two time scales $\mu^{-1}$ and $D_R^{-1}$ leads to fascinating dynamical features as we 
demonstrate below. The physical picture emerging from our study is summarized in Fig.~\ref{fig:snap1} for $D_R^{-1} < \mu^{-1}$ 
(upper panel) and for $D_R^{-1} > \mu^{-1}$ (lower panel). In both cases, at short-times $t \ll \min(D_R^{-1}, \mu^{-1})$, the 
presence of the activity gives rise to strong anisotropy with the particle retaining its initial orientation (chosen to be along 
$x$-direction here). In this regime, the effect of the trap can be neglected and the dynamics reduces to that of a free ABP. At 
later times, if $D_R^{-1} < \mu^{-1},$ the anisotropy starts to disappear and the ABP undergoes ordinary diffusion (upper middle 
panel). Eventually, for $t \gg \mu^{-1}$ the probability distribution saturates to a Boltzmann-like form with a single Gaussian 
peak at the center of the trap.  On the other hand, for strongly active system, \ie, when $D_R^{-1} > \mu^{-1}$ the anisotropy 
persists and the particle starts to accumulate away from the center of the trap. For $t \gg D_R^{-1}$ the isotropy is slowly 
recovered (lower right panel). The stationary distributions we obtain in the two limiting cases are in agreement with the 
experimental and numerical observations~\cite{Takatori,Solon2015}.

 \begin{figure*}[t]
\includegraphics[width = \linewidth]{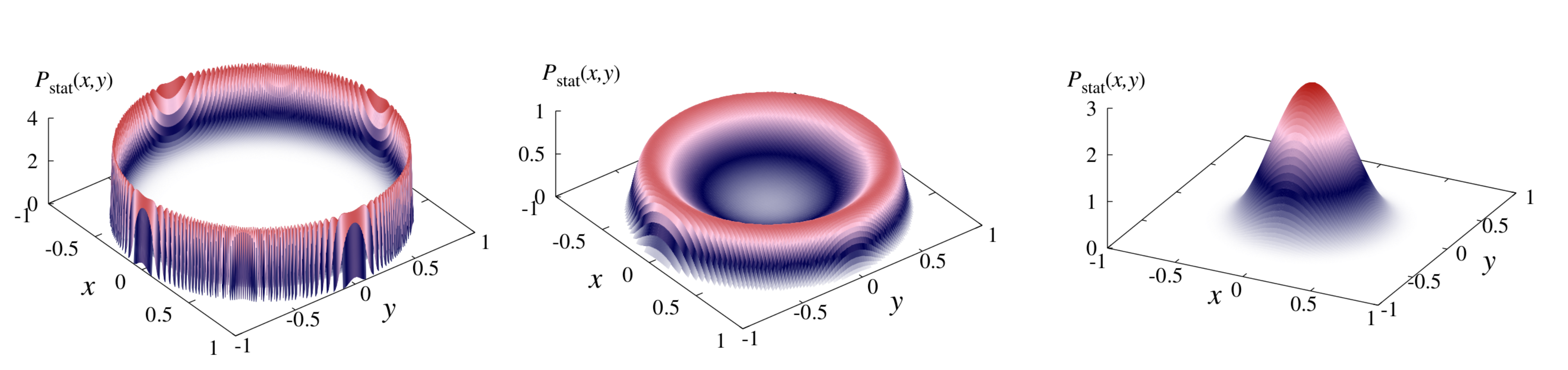}
 \caption{Stationary distribution $P_{\textrm{stat}}(x,y)$ of an ABP in a harmonic trap for different values of $D_R=0.1$ (left), $D_R=1.0$ (centre) and $D_R=10.0$ (right). The left and middle panel show the delocalized state where the particle is most likely to be accumulated away from the center. The right panel corresponds to the passive limit where the stationary distribution is Gaussian.  Here the trap stiffness $\mu=1.0$ and $v_0=1.0.$}\label{fig:Pxy}
\end{figure*}

The position distribution $P_\mu(x,y,t)$ can be obtained by integrating out the orientational degree of freedom from the full 
probability density ${\cal P}_\mu(x,y,\phi,t)$ :
\bea
P_\mu(x,y,t) = \int \id \phi ~{\cal P}_\mu(x,y,\phi,t) \;.
\eea
Starting from the Langevin equations (\ref{eq:trap_modely}), it is easy to write down the corresponding Fokker-Planck equation,
\bea\label{FP_SM1}
\partial_t {\cal P}_\mu(x,y,\phi, t) &=& \frac{\partial}{\partial x} \bigg [(\mu x - v_0 \cos \phi) {\cal P}_\mu\bigg] \cr
& +& \frac{\partial}{\partial y} \bigg[(\mu y - v_0 \sin \phi) {\cal P}_\mu\bigg] + D_R \frac{\partial^2 {\cal P}_\mu}{\partial \phi^2},\qquad \label{eq:FP_trap}
\eea
where we have suppressed the argument of ${\cal P}_\mu$ on the right hand side for brevity. 

In the long time limit the position distribution $P_\mu(x,y,t)$ converges to a stationary form  which is denoted by 
\bea\label{Pstat}
P_{\textrm{stat}}(x,y) = P_\mu(x,y,t \to \infty) \;.
\eea
 
Unfortunately, the Fokker-Planck equation \eqref{eq:FP_trap} is hard to solve, even for the stationary state.  Very recently, in 
Ref.~\cite{Malakar2019}, the same Langevin equation (\ref{eq:trap_modely}) was studied, but in the presence of an additive 
translational noise in the $x$ and $y$ directions with a nonzero translational diffusion constant $D_T$. The stationary 
distribution $P_{\textrm{stat}}(x,y)$ was computed from the associated Fokker-Planck equation as a power series expansion in 
terms of the parameter $\lambda = \frac{v_0}{\sqrt{D_R D_T}}.$ However, this result cannot be easily extrapolated to the case 
$D_T=0$ where $\lambda\to \infty$ (except in the strongly passive case where $D_R\to \infty$ 
limit is taken first). 
This is because, in general, the two limits do not commute: (i) first $D_T \to 0$ and then $t\to \infty$ (ii) first $t\to \infty$ 
with finite $D_T$ and then $D_T \to 0$. While we are interested in limit (i), Ref.~\cite{Malakar2019} studied mostly the limit 
(ii).

Here we follow a different approach that involves deriving and solving an exact recursion 
relation satisfied by the moments of the 
position. A similar method involving recursion of moments was studied by Gredat, Dornic and Luck (GDL) in Ref. \cite{Gredat} in 
the context of a reaction diffusion equation. In their problem, GDL were interested in the (imaginary) exponential functional of 
a Brownian motion with a nonzero drift.  Here we adapt their approach to our ABP problem in a harmonic trap. Our recursion 
relation, though formally appears deceptively similar to that of GDL, the slight difference actually leads to very different 
physics and results. Indeed in Appendix \ref{sec:GDL} we will discuss in detail the differences between the two recursion 
relations.
 
It is first convenient to recast the Langevin equations \eqref{eq:trap_modely} in terms of a complex coordinate $z=x+iy.$ Our 
goal is to evaluate the moment of the type
\bea\label{def_Mkl}
M_{k,l}(t) = \la z^k(t) \bar z^l(t) \ra, \;
\eea
where $\bar z(t) = x(t) - i y(t) $ is the complex conjugate of $z.$ From Eq.~ \eqref{eq:trap_modely} it immediately follows that 
$z(t)$ evolves according to,
\bea
\dot z = -\mu z + v_0 e^{i \phi(t)}. \label{eq:zdot_trap}
\eea
which can be formally solved to get,
\bea
z(t) = v_0 \int_0^t \id s~ e^{-\mu(t-s)} e^{i\phi(s)}\label{eq:zt_trap}
\eea
We assume that the particle starts initially at the origin $x=y=0$ with $\phi(0) =0$. In principle, one can use 
Eq. (\ref{eq:zt_trap}) and the Gaussian property of the process $\phi(s)$,
to express $M_{k,l}(t)$ as a $(k+l)$-fold multiple integral. However, evaluating 
this multiple integral explicitly seems very hard. Instead, we will derive below an exact recursion relation for the moments 
$M_{k,l}(t)$.

To proceed further, it is useful to discretize the continuous time expression of Eq.  (\ref{eq:zt_trap}) in a discrete-time 
setting. We imagine that the interval $[0,t]$ consists of $n$ discrete intervals each of length $\ve>0$, such that $t = n \ve$. We 
then split the time interval $[0,t]$ in the integral in Eq. (\ref{eq:zt_trap}) into two separate intervals $[0,\ve]$ and $[\ve, 
t]$. This gives
\begin{equation}\label{split1}
z(t) = v_0 \left(\int_0^\ve e^{-\mu(t-s)+ i \phi(s)}\, ds + \int_\ve^t e^{-\mu(t-s)+ i \phi(s)}\, ds \right) \;.
\end{equation} 
The first integral, to leading order in $\ve$, gives $e^{-\mu \,t}\ve$, where we used $\phi(0)=0$. In the second integral, we 
make a change of variable $s = \ve + \tau$ and rewrite it as, $\int_0^{t-\ve} e^{-\mu(t-\ve-\tau) + i \phi(\ve+\tau)} d\tau$. 
Next we write $\phi(\ve+\tau) = \phi(\ve+\tau) - \phi(\ve) + \phi(\ve)$, \ie, add and subtract $\phi(\ve)$. Putting this 
together, we get
\begin{equation} \label{split2}
z(t) \approx v_0 \left( e^{-\mu t} \ve + e^{i \phi(\ve)}\int_0^{t-\ve} e^{-\mu(t-\ve-\tau) + i \tilde \phi(\tau)} d\tau\right) \;,
\end{equation}
where 
\bea\label{phitilde}
\tilde \phi(\tau) = \phi(\ve+\tau) - \phi(\ve) \;.
\eea
Now we will use the crucial property that $\tilde \phi(\tau)$ is also a Brownian motion starting at $\tilde \phi(0)=0$, and with 
correlation function $\la \tilde \phi(t_1) \tilde \phi(t_2) \ra = 2 D_R\, \min(t_1,t_2)$. Importantly, the statistical properties 
of $\tilde \phi(t)$ do not depend on $\ve$. In other words, one can write a statistical identity in law
\bea\label{id_law}
\tilde \phi(\tau) \equiv \phi(\tau) \;,
\eea 
where $\equiv$ means that the right hand side and left hand side have identical distributions. Consequently, using this identity 
(\ref{id_law}) and the definition of $z(t)$ in Eq. (\ref{eq:zt_trap}), one gets
\bea\label{id_int}
\int_0^{t-\ve} e^{-\mu(t-\ve-\tau) + i \tilde \phi(\tau)} d\tau \equiv z(t-\ve) \;.
\eea
Hence, (\ref{split2}) provides us with a statistical identity
\bea\label{split3}
z(t) \equiv v_0 \, \ve \, e^{- \mu t} + e^{i \phi(\ve)} z(t-\ve) \;.
\eea
Denoting $z_n = z(t=n\ve)$ in the discrete-time setting, we then obtain a Kesten type statistical recursion relation  
\bea\label{kesten}
z_n \equiv v_0 \; \ve \; e^{-\mu n \ve} + \eta_n \; z_{n-1} 
\eea
where $\eta_n = e^{i\phi(\ve)}$ is an effective noise, independent of $z_{n-1}$. The complex conjugate $\bar z_n$ also satisfies 
a similar relation,
\bea\label{kesten_conj}
\bar z_n = v_0 \; \ve \; e^{-\mu \ve} + \bar \eta_n \; \bar z_{n-1} \;,
\eea
where $\bar \eta_n$ is the complex conjugate of $\eta_n.$ 
Using the Gaussian property of $\phi(s)$, %(\ref{identity_Gauss}),
one can easily evaluate the moments of the noise $\eta_n$. For instance, one gets $\la \eta_n\ra = e^{-\ve D_R}$ and correlation 
$\la \eta_n^k ~\bar \eta_n^l \ra = e^{-\ve D_R(k-l)^2}.$

Using Equations \eqref{kesten} and \eqref{kesten_conj} one can now derive a recursion relation for the discrete-time version of 
the moment $M_{k,l}(n) = \la z_n^k \bar z_n^l \ra$. We take $z_n^k$ in Eq. (\ref{kesten}) and $\bar z_n^l$ in Eq. 
(\ref{kesten_conj}), multiply them and then take the expectation value with respect to the noise $\eta_n$. We use the 
independence of $\eta_n$ and $z_{n-1}$ and the known moments of the noise $\eta_n$ and then expand in powers of $\ve$. Keeping 
terms only up to order $O(\ve)$, we get
\bea
M_{k,l}(n) &\simeq& [1- \ve D_R (k-l)^2] M_{k,l}(n-1) \n \\
&& + v_0 \ve e^{-\mu n \ve} [k M_{k-1,l}(n-1) + l M_{k,l-1}(n-1)] \;. \n \\
\eea
Taking the continuous-time limit $\ve \to 0$ and replacing $(M_{k,l}(n)-M_{k,l}(n-1))/\ve$ by the time derivative $dM_{k,l}/dt$ 
we arrive at the exact recursion relation
\bea
\dot M_{k,l} &=&-D_R(k-l)^2 M_{k,l} + v_0 e^{-\mu t}[k M_{k-1,l}+ lM_{k,l-1}] \label{eq:Mkl} \cr
&& 
\eea
with the conditions $M_{0,0}(t) = 1$ at all times and $M_{k,l}(0)=0$ for $k,l>0$. We also use the convention $M_{k,l}(t) = 0$ for 
$k,l<0$. It is easy to check that $M_{k,l}(t) = M_{l,k}(t)$. Eq.~\eqref{eq:Mkl} allows us to compute the moments explicitly in a 
recursive fashion (see Appendix \ref{sec:Mkl} for the first few values of $k,l$).

Note that, since the right hand side is explicitly time-dependent, it is not possible to obtain the stationary state by simply 
equating $\dot M_{k,l}$ to zero, rather one has to find the full time-dependent solution and then take long-time limit to find 
the same. It turns out that this can be done in the two limiting cases, $D_R \to \infty$ (strongly passive) and $D_R \to 0$ (strongly 
active) which are discussed in details below. \\

\noindent {\it Strongly passive limit} ($D_R \to \infty$): To solve the moment evolution Eq.~\eqref{eq:Mkl} in the limit of $D_R \to \infty$ we inspect the large $D_R$ behaviour of the first few moments presented in Eq.~\eqref{eq:M012} in Appendix \ref{sec:Mkl}. It turns out that these quantities, to the leading order in $D_R^{-1},$ are of the form,
\begin{equation}
M_{k,l}(t) \simeq \frac{v_0^{k+l} k! }{[(k-l)!]^2} \left[\frac{e^{-\mu t}}{D_R} \right]^k  \left[\frac{e^{\mu t} -e^{-\mu t}}{\mu}\right]^l, \; k \ge l \;.
\;\;\;\label{eq:Mkl_passive}
\end{equation}
Indeed, substituting this ansatz in the recursion relation (\ref{eq:Mkl}), it can be verified that Eq.~\eqref{eq:Mkl} is indeed satisfied by Eq.~\eqref{eq:Mkl_passive}, up to leading order for large $D_R$. Note that this leading order result for $M_{k,l}(t)$ in Eq. (\ref{eq:Mkl_passive}) is actually valid for all time $t$, including $t=0$. 

To extract further information, we consider the diagonal moments $M_{k,k}(t) = \la (z(t) \bar z(t))^k\ra$. Using $z(t) \bar z(t) = x^2(t) + y^2(t) = r^2(t)$, the diagonal element $M_{k,k}(t) = \la r^{2k}(t)\ra$ is precisely the $2k$-th radial moment of the full distribution. This radial moment is given by,
\bea\label{rad_moment}
\la r^{2k}(t)\ra = \int_0^{\infty} r^{2k+1}\, P_{\textrm{rad}}(r,t) \, dr \;.
\eea
where $P_{\textrm{rad}}(r,t)$ is the marginal radial distribution,
\bea\label{Prad}
P_{\textrm{rad}}(r,t) = \int_0^{2 \pi} P_\mu(r,\theta,t) \; d\theta \;.
\eea
Here $P_\mu(r,\theta,t)$ denotes the position probability in the radial coordinates, and is equivalent to 
$P_\mu(x,y,t)$~\cite{footnote1}. Setting $l=k$ in Eq. (\ref{eq:Mkl_passive}) we then 
get
\begin{equation}
\la r^{2k}(t) \ra = M_{k,k}(t) \simeq \Gamma(k+1) \left[\frac{v_0^2}{\mu D_R}\left(1-e^{-2\mu t} \right)\right]^k. \label{eq:r2k_Dinf}
\end{equation}
Anticipating a Gaussian behaviour for the radial distribution, we make the ansatz, and check a posteriori, that $P_{\textrm{rad}}(r,t)$ has the form $P_{\textrm{rad}}(r,t) = A(t)\, e^{-B(t) r^2}$. Substituting this ansatz in Eq. (\ref{rad_moment}) and comparing to the result in (\ref{eq:Mkl_passive}), we see that 
\bea\label{AB}
A(t) = 2 B(t) \;\;, \;\; B(t) = \frac{\mu D_R}{v_0^2(1-e^{-2\mu t})} \;.
\eea
Finally, this gives
\bea\label{Prad_final}
P_\text{rad}(r,t) \simeq \frac{2\,\mu D_R }{v_0^2(1-e^{-2\mu t})}\exp{\left[-\frac{\mu D_R r^2 }{v_0^2(1-e^{-2\mu t})} \right]}.
\eea

Note that this solution is valid at all times $t$. In particular, at early times, when $D_R^{-1} \ll t \ll \mu^{-1}$ the solution in Eq.~(\ref{Prad_final}) corresponds to free isotropic diffusion with a diffusion constant $D_\text{eff}=v_0^2/2 D_R$. 
This scenario  corresponds to the upper middle panel in Fig.~\ref{fig:snap1}. In contrast, when $t \gg \mu^{-1}$, the radial distribution (\ref{Prad_final}) approaches a stationary form.

Moreover, from Eq.~\eqref{eq:Mkl_passive} it follows that for $k \ne l,$ $M_{k,l}(t)$ decays exponentially with time and vanishes in the long time limit. This indicates that the distribution quickly loses the anisotropy and the stationary distribution becomes radially symmetric. Consequently, the stationary position distribution in Eq. (\ref{Pstat}) is given by
\bea\label{Pstat_2}
P_{\textrm{stat}}(x,y) &=& \frac{1}{2 \pi} P_{\textrm{rad}}(r,t\to \infty) \;.
\eea
Using Eq. (\ref{Prad_final}), one gets the expected Boltzmann distribution
\bea\label{Pstat3}
P_{\textrm{stat}}(x,y) =  \frac{\mu D_R }{\pi v_0^2}\exp{\left[-\frac{\mu D_R (x^2+y^2) }{v_0^2} \right]},
\eea
with an effective temperature $T_{\textrm{eff}} =  v_0^2/2 D_R = D_\text{eff}$, in full agreement with the experimental observation~\cite{Takatori}. \\

\noindent{\it Strongly active limit ($D_R = 0$)}: In this case, the first term on the right hand side of Eq.~\eqref{eq:Mkl} drops out and it can be checked that
\bea\label{Mkl_active}
M_{k,l}(t) = \left[\frac {v_0} \mu \left(1- e^{-\mu t} \right) \right]^{k+l}
\eea
solves the resulting equation at all times $t$. Again, setting $l=k$ in (\ref{Mkl_active}) 
the time-dependent radial moments are given by
\bea
\la r^{2k}(t) \ra = M_{k,k}(t) = \left[\frac{v_0}\mu (1- e^{-\mu t}) \right]^{2k}. \label{eq:r2k_D0}
\eea
Comparing Eq. (\ref{rad_moment}) with Eq. (\ref{eq:r2k_D0}) gives the time-dependent marginal radial distribution,
\bea
P_{\textrm{rad}}(r,t) =  \frac {\mu} {v_0 (1-e^{-\mu t}) } \delta \left[r-\frac{v_0(1-e^{-\mu t})}{\mu} \right]\;. \label{eq:Prt_D0}
\eea
Note however that strictly for $D_R=0$, the position distribution $P(x,y,t)$ is not radially symmetric. Indeed, in this case, the Langevin equation (1) in the main text reduces to a pair of deterministic equations:
\bea\label{Langevin_DR0}
&&\dot x = - \mu x + v_0 \;\; {\textrm{and}} \;\; \dot y = - \mu y \;,
\eea 
with initial conditions $x(0) = y(0) = 0$. Solving these equations give $x(t) = (v_0/\mu)(1-e^{-\mu \,t})$ and $y(t) = 0$. Consequently, the position distribution function is given~by 
\bea\label{Pxy_active}
P_\mu(x,y,t) = \delta\left (x - \frac{v_0(1-e^{-\mu t})}{\mu}\right) \delta(y) \;.
\eea
One can check that the moment $M_{k,l}(t)$ computed with this distribution is indeed given by (\ref{Mkl_active}). Moreover, the radial marginal distribution $P_{\textrm{rad}}(r,t)$ computed from this two-dimensional distribution is indeed given by (\ref{eq:Prt_D0}). 

Thus strictly for $D_R = 0$ the position distribution in the $2d$-plane is highly anisotropic. This is true even in the $t \to \infty$ limit, where we see from Eq. (\ref{Mkl_active}) that 
\bea\label{Mkl_inf}
M_{k,l}(t \to \infty) = \left( \frac{v_0}{\mu}\right)^{k+l}  \;\; {\textrm{for  all}} \;\; k,l \;.
\eea
Thus, the off-diagonal elements remain non-zero as $t \to \infty$, indicating the presence of anisotropy in the stationary state.

However, for any finite $D_R > 0$, the rotational diffusion spreads the particle position uniformly over the angle $[0,2 \pi]$. Consequently, in the long time limit and $D_R \to 0^+$, the position distribution approaches a stationary form that is fully isotropic in the $2d$ plane. Indeed, from the exact expression for the moments in (\ref{eq:M012}), it is easy to verify that, for $D_R \to 0^+$, the off-diagonal elements decay as
$M_{k,l}(t) \sim e^{- D_R(k-l)^2\,t}$ at late times, for $k \neq l$. In particular, for $t \gg D_R^{-1}$, $M_{k,l}(t) \to 0$ for $k \neq l$. In contrast, the diagonal elements approach to non-zero values as $t \to \infty$. More precisely, we find
\bea\label{DRneq0}
&&M_{k,k}(t \to \infty) \to \left(\frac{v_0}{\mu}\right)^{2k} \cr
&& M_{k,l}(t \to \infty) \to 0 \;, \;\;\;\;\;  k \neq l. 
\eea
Note the difference with the strictly $D_R = 0$ case in Eq.~(\ref{Mkl_inf}). Consequently, in this $D_R \to 0^+$ limit, for $t \gg D_R^{-1}$, it follows from Eq.~(\ref{DRneq0}) that the position distribution approaches an isotropic form in the stationary limit and is given by 
\bea
P_{\textrm{stat}}(x,y) = \frac {\mu} {2 \pi v_0} \delta \left[\sqrt{x^2+y^2}-\frac{v_0}{\mu} \right] \label{eq:Pr_active}
\eea
where the particle is strongly confined at the boundary of the trap $r_b=v_0/\mu$. This non-Boltzmann distribution results from the strongly active nature of the dynamics.

Figure \ref{fig:Pxy} shows the stationary distribution $P_{\textrm{stat}}(x,y)$ in the $(x,y)$ plane obtained from simulations, for different  $D_R.$ As $D_R$ decreases, the stationary distribution shows a crossover from the passive regime, with a single-peaked Gaussian around $r=0$, to the active regime, with a delocalized state where the particle is confined around a narrow ring away from the origin, at $r_b = v_0/\mu$.

\section{Conclusion} \label{conclusion}

To summarize, this paper has two parts. In the first part, we have studied the late time 
dynamics of a free ABP in two dimensions, focussing on the position distribution $P(x,y,t)$. 
We have showed that while the typical fluctuations are described by a Gaussian distribution 
as expected from the central limit theorem, large fluctuations, where $\sqrt{x^2+y^2} \sim 
\cal O(v_0 t)$, are described by non-Gaussian tails. These rare fluctuations capture the 
signature of `activity' even at late times $t$. In this regime we have showed that 
$P(x,y,t)$ admits a large deviation form $P(x,y,t) \sim \exp\left [ -D_R t \Phi(z) \right]$ 
where $z=\sqrt{x^2+y^2}/(v_0 t)$. We have computed the rate function $\Phi(z)$ both 
analytically and numerically.

Another way to observe the fingerprints of activity in the position distribution at late 
times is to switch on an external harmonic potential with stiffness $\mu$. In this case the 
position distribution approaches a stationary form at late times and the stationary 
distribution $P_{\text{stat}}(x,y)$ depends explicitly on the activity parameter $D_R^{-1}$. 
We compute the stationary distribution explicitly in the two opposite limits: (i) strongly 
active ($D_R\to 0$) and (ii) strongly passive ($D_R\to \infty$). In the former case the 
distribution is ring shaped with ring radius $r= v_0/\mu$, while in the latter case it is a 
Gaussian centered at the origin. As $D_R$ increases the shape of the distribution smoothly 
crosses over from the ring shape to the Gaussian shape. This is in agreement with the 
results seen in experiments ~\cite{Takatori} and simulations~\cite{Solon2015,Potosky2012}.

We find it remarkable that even for this simplest ABP model (free or harmonically confined) 
the position distribution $P(x,y,t)$ cannot be computed exactly at all times in the real 
space. At least in this paper we managed to compute analytically the large deviation 
function that describes the atypical fluctuations at late times for the free ABP. Of course 
there are many interesting open questions related to our work. For example, it would be 
interesting to study the dynamics of an ABP in higher dimensions and derive the associated 
rate function $\Phi(z)$. In this paper we have focused on a single ABP---it would be 
interesting to derive the large deviation function associated with the late time density 
profile of a gas of {\it interacting} ABP's. Finally, in the presence of a confining 
potential, we have studied the stationary state in the case of an isotropic harmonic trap. 
It would be interesting to study the position distribution of an ABP in an anisotropic 
harmonic trap, or more generally for anharmonic traps, in two or higher dimensions.

\acknowledgements
We thank I. Dornic and J. M. Luck for useful discussions and for pointing out the Ref.~\cite{Mumford} to us.
We acknowledge support from the project 5604-2 of the Indo-French Centre for the 
Promotion of Advanced Research (IFCPAR). SNM acknowledges 
the support from the Science
and Engineering Research Board (SERB, government
of India) under the VAJRA faculty scheme (Ref.
VJR/2017/000110) during a visit to the Raman Research Institute in 2019,
where part of this work was carried out. U.B. acknowledges support from Science and Engineering Research Board (SERB), India under Ramanujan Fellowship (Grant No. SB/S2/RJN-077/2018).

\appendix

\section{Exact solution of Mathieu Eigenfunctions}\label{sec:a0}

As explained in Section \ref{sec:LDF}, we are interested only in the $\pi$-periodic even 
solutions of the Mathieu equation
\bea
\text{ce}_{2n}''(v,q) + (a_{2n}(q)- 2 q \cos 2 v)\text{ce}_{2n}(v,q)= 0 
\eea
where $a_{2n}(q)$ are the associated eigenvalues.
To calculate the moments and the large deviation function 
we only need the lowest eigenvalue. The series expansion of that 
lowest eigenvalue $a_0(q)$ is known for both in the small $q$ and large $q$ limit. 
For small $q,$  
\bea
a_0(q) = \sum_{n=1}^{\infty} \alpha_{2n} q^{2n}. \label{eq:a0_small}
\eea
The first few coefficients are quoted here,
\bea
\alpha_2 &=& -\frac 12,\qquad \alpha_4 = \frac 7{128}, \cr
\alpha_6 &=& - \frac{29}{2304}, ~~\alpha_8 = \frac{68687}{18874368}, \;\; \cdots \label{eq:alpha_n}
\eea
On the other hand, in the large $q$ limit, the expansion is given by,
\bea
a_0(q) = \sum_{n=0}^\infty \beta_n q^{1-\frac n2}\label{eq:a0_large}
\eea
where
\bea
\beta_0 &=& -2, \;\; \beta_1 = 2, \;\;\beta_2 = -\frac 14,\cr
\beta_3 &=& -\frac 1{32}, ~\beta_4 = - \frac 3{256}, 
\beta_5=-\frac {53}{8192}\; \cdots \label{eq:beta_n}
\eea

% \bea
% a_0(q) \simeq -2 q +2 \sqrt{q} - \frac 14 - \frac 1{32 \sqrt{q}} - \frac 3{2 56 q} + \cdots
% \eea

\section{Systematic determination of $\Phi(z)$}\label{sec:A_Phiz}

Equation \eqref{eq:Phiz} in the main text relates the large deviation function 
$\Phi(z)$ to the eigenvalue $a_0$ through a Legendre transform,
\bea
\min_{-1 \le z \le 1} \bigg[\frac p{D_R}z + \Phi(z)\bigg] = 
\frac 14 a_0\left(\frac{2p}{D_R}\right)
\eea
The large deviation function can be extracted from the inverse transform,
\bea
\Phi(z) &= & \max_{h \in \mathbb{R}} \bigg[\frac 14 a_0(2h) - hz \bigg]
\eea
where we have defined $h=p/D_R.$  The large deviation function is then given by,
\bea
\Phi(z) = \frac 14 a_0(2h^*(z)) - z\, h^*(z) \label{eq:Phi_hstar}
\eea
where $h^*(z)$ is the value of $h$ corresponding to the maximum of the function 
$g_z(h)=\frac{1}{4}\, a_0(2h) - z\,h,$  
and can be obtained by setting its derivative to zero, \ie, by solving 
\bea
\frac{1}{4} \frac{\id }{\id h} a_0(2h) = z. \label{eq:hmax}
\eea
As $a_0(2h)$ is known as the sum of an infinite series in $h$ 
(see Eqs.~\eqref{eq:a0_small}-\eqref{eq:a0_large}), it is best to solve the above 
equation recursively. 
It is easy to see that for small values of $z,$ $h^*$ is also small 
while for $z \to \pm 1,$ the maximum occurs at large values of 
$h^*$~\cite{footnote2}.
It is then convenient to use Eq.~\eqref{eq:a0_small} (respectively 
Eq.~\eqref{eq:a0_large}) for finding $h^*(z)$ near $z=0$ (respectively near $z = \pm 
1$).

% In the following we 
%illustrate how the function $\Phi(z)$ can 
%systematically be obtained as a power series in 
%$z,$ by solving the above equation recursively and using the series expansion of $a_0(2h).$

Let us first look at the case $z \approx 0.$ In this case, 
using the series \eqref{eq:a0_small}, Eq.~ \eqref{eq:hmax} becomes,
\bea
\frac 12 \sum_{n=1}^\infty \alpha_{2n} n 2^{2n} (h^*)^{2n-1} =z \label{eq:h*}
\eea
In the following we solve this equation recursively to 
systematically determine $\Phi(z)$ as a series in $z.$ 
To the lowest order, \ie, keeping the term linear in $h$ only, we have, 
\bea
2 \alpha_2 h^* =z,
\eea
which, using the value of $\alpha_2$ 
(see Eq.~\eqref{eq:alpha_n}) yields $h^*= -z.$ This value of $h^*$, 
substituted in Eq. \eqref{eq:Phi_hstar}, and keeping the lowest order term again, gives,
\bea
\Phi(z) \approx \frac 12 z^2. \label{eq:Phiz1}
\eea
Equation \eqref{eq:Phiz1} implies that, close to the origin $z=0,$ 
in the long time limit, the position distribution is Gaussian. 
The higher order corrections can also be systematically calculated in a recursive manner. 

Since both $\Phi(z)$ and $a_0(2h)$ are even functions of their arguments, 
it is easy to see that $h^*$ must be an odd function of $z,$ and 
we can write a series expansion,
\bea
h^*(z) = \sum_{m=1,3,\cdots}^\infty c_m z^m.\label{eq:sol_h*}
\eea
Substituting this form in Eq.~\eqref{eq:h*}, 
and then comparing coefficients of powers of $z$ on both sides, 
one can solve for the  $c_m$ recursively. 
Clearly, $c_1=-1,$ as we have explicitly shown above. 
The next few coefficients are computed using Mathematica and  are quoted below,
\bea
c_3 = -\frac 78, c_5 =-\frac {209}{192}, c_7 = - \frac{53231}{294912}
\eea
Using these coefficients, and substituting Eq.~\eqref{eq:sol_h*} in Eq.~\eqref{eq:Phi_hstar} one can construct $\Phi(z)$ 
as a series expansion in $z,$ which is given in Eq. \eqref{eq:Phiz_pm1}
 in the main text.

The behaviour of $\Phi(z)$ near the boundaries $z = \pm 1$ can also be extracted in a 
similar manner. As $\Phi(z)$ is an even function of $z,$ it suffices to compute it near one 
boundary, say $z=-1.$ We follow the same procedure as traced above, but use 
Eq.~\eqref{eq:a0_large} for $a_0.$ Accordingly, Eq.~\eqref{eq:h*} becomes,
\bea
\sum_{n=0}^\infty \frac{\beta_n}{2^{1+\frac n2}}\big(1-\frac n2\big) (h^*)^{-\frac n2} =z
\eea
which we solve order by order to find $h^*(z).$

To the lowest order, we have,
\bea
\beta_0 + \frac{\beta_1}{2^{3/2}\sqrt{h^*}} = 2 z \label{eq:h*_z1}
\eea
which, after substituting the values of $\beta_0$ and $\beta_1,$ yields, $h^* = 1/8(1+z)^2.$ 
Using this value of $h^*$ in Eq.~\eqref{eq:Phi_hstar}, we get, near $z=-1,$
\bea
 \Phi(z) \approx \frac 1{8(1+z)}.
\eea
The higher order corrections are systematically obtained by assuming a 
series expansion for $h^*,$
\bea
h^*(z) = \sum_{n=-2}^{\infty} b_n (1+z)^n \label{eq:sol_hz1}
\eea
where $b_{-2}=1/8,$ as shown above. The coeffcients $b_n$ for $n>-2$ can be obtained by substituting Eq.~\eqref{eq:sol_hz1} in Eq.~\eqref{eq:h*_z1} and equating coefficients of powers of $1+z$ on both sides. This exercise gives,
\bea
b_{-1} = 0,~ b_0 = \frac 1{64}, ~b_1 = \frac 3{128}.
\eea
The large deviation function $\Phi(z)$ near $z=-1$ 
is then obtained using Eq.~\eqref{eq:sol_hz1} in Eq.~\eqref{eq:Phi_hstar}, and is given by,
\bea
\Phi(z)&=&\frac 1{8(1+z)} -\frac{1}{16} - \frac{(1+z)}{64}  \cr
&-& \frac{3}{256} (1+z)^2 - \frac{51}{4096} (1+z)^3  + \cdots. 
\eea
Using the symmetry of $\Phi(z),$ its behaviour near $z=1$ can be obtained from the above equation by substituing $z \to -z.$ This is quoted in the main text in the second line of  Eq.~\eqref{eq:Phi_boundary}.

\section{Connection to GDL}\label{sec:GDL}

In Ref. \cite{Gredat} Gredat, Dornic, Luck (GDL)  were interested in the imaginary exponential functional of a Brownian motion and studied an effective process given by
\bea\label{z_GDL}
z^{GDL}(t) = v_0 \int_0^t e^{-\mu s + i \phi(s)} \,ds  \;.
\eea
The two processes, $z(t)$ in (\ref{eq:zt_trap}) and $z^{GDL}(t)$ in (\ref{z_GDL}), look deceptively similar. However it turns out that they have rather different properties and in fact the recursion relation for the moments turn out to be rather different. 

A recursion relation for the discretized version of $z^{GDL}(t)$ can be derived following scheme similar to the one used for $z(t)$ in the main text, and yields \cite{Gredat},
\bea\label{recursion_GDL}
z_n^{GDL} \equiv v_0 \, \ve + e^{-\mu \ve} \, \eta_n \, z_{n-1}^{GDL} \;,
\eea
which is manifestly different from our recursion relation~(\ref{kesten}). 

Correspondingly, the recursion relation for the moments $\tilde M_{k,l}(t) =  M^{GDL}_{k,l}(t)$ in the GDL case  also turns out to be very different \cite{Gredat},
\bea\label{Mkl_GDL}
\frac{d}{dt}{\tilde M_{k,l}} &=& -\left(\mu(k+l) + D_R(k-l)^2 \right)\,{\tilde M_{k,l}} \n \\
&+& v_0\left(k\, {\tilde M_{k-1,l}} + l \,{\tilde M_{k,l-1}} \right) \;.
\eea
Note that there is no explicit time dependence on the right hand side of this equation (\ref{recursion_GDL}) and the moments in the stationary state can be simply obtained by setting the time derivative to be zero on the left hand side of (\ref{recursion_GDL}). As mentioned above, the situation in our case is completely different.

\section{Solution of the moment recursion relation}\label{sec:Mkl}
The moments $M_{k,l}(t)$ evolve according to,
\bea
\dot M_{k,l} &=&-D_R(k-l)^2 M_{k,l} + v_0 e^{-\mu t}[k M_{k-1,l}+ lM_{k,l-1}]  \cr
&& \label{eq:dot_Mkl}
\eea
We can think of $(k,l)$ as the grid points on the $2d$ lattice with $k,l \geq 0$. We note that by definition $M_{0,0}(t) = 1$ at all times $t$. As a result, it is easy to see from the recursion relation (\ref{eq:dot_Mkl}) that the solution $M_{k,l}(t)$ is symmetric under exchange of $k$ and $l$, \ie, 
\bea\label{symmetry}
M_{k,l}(t) = M_{l,k}(t) \;.
\eea
Hence, it is sufficient to study $M_{k,l}(t)$ only for $k \geq l$. The recursion relations for the first few values of $k$ and $l$ read, for instance (with the convention that $M_{k,l}(t) = 0$ for $k,l<0$)  
\bea
\dot M_{1,0}(t) &=& - D_R M_{1,0}(t) + v_0 e^{-\mu t} M_{0,0}(t) \cr
\dot M_{1,1}(t) &=&  2 v_0 e^{-\mu t}  M_{1,0}(t)  \cr
\dot M_{2,0}(t) &=& -4 D_R M_{2,0}(t) + 2 v_0 e^{-\mu t} M_{1,0}(t) 
%\dot M_{21}(t) &=& -2 D_R M_{21}(t) +  v_0 e^{-\mu t} [2 M_{11}(t)+ M_{20}(t)] \cr
\label{eq:dot_M012}
\eea
and so on.  These equations can be solved recursively, \ie, using the solution of the previous equation. The solution of these first few moments can be written explicitly at all times $t$,
\begin{widetext}
\bea
M_{1,0}(t) &=&  \frac{v_0 (e^{-\mu t} - e^{-D_R t})}{D_R - \mu}  \cr
M_{1,1}(t) &=&  \frac{v_0^2} {(D_R -\mu)} \left[\frac{1-e^{-2 \mu t}}{\mu} -  \frac{2(1-e^{-(D_R+\mu)t})}{D_R + \mu}\right] \n \\[0.5em]
M_{2,0}(t) &=& \frac{v_0^2 [(3D_R -\mu)e^{-2\mu t}-2(2D_R -\mu)e^{-(D_R+ \mu)t}+(D_R -\mu)e^{-4D_Rt}]}{(D_R -\mu)(2D_R -\mu)(3D_R -\mu)} \;.\label{eq:M012}
\eea
\end{widetext}
As we see, the solutions quickly become long and cumbersome as $k$ and $l$ increase. Fortunately, Eq.~\eqref{eq:dot_Mkl} can be solved exactly to find $M_{k,l}(t)$ for all $k$ and $l$ in the two limiting cases ($D_R \to \infty$ and $D_R = 0$) which is done in the main text.

%To extract more specific informations, we now investigate two limiting cases ($D_R \to \infty$ and $D_R = 0$) where $M_{k,l}(t)$ can be obtained explicitly for all $k$ and $l$.   


\begin{thebibliography}{99}

\bibitem{Romanczuk} P. Romanczuk, M. B\"{a}r, W. Ebeling, B. Lindner, and 
L. Schimansky-Geier, Eur. Phys. J. Special Topics {\bf 202}, 1 (2012).
% Active brownian particles

\bibitem{soft} M. C. Marchetti, J. F. Joanny, S. Ramaswamy, T. B. Liverpool, J. Prost, M. Rao, and R. Aditi Simha, Rev. Mod. Phys. {\bf 85}, 1143  (2013).
% Hydrodynamics of soft active matter

\bibitem{BechingerRev} C. Bechinger, R. Di Leonardo, H. L\"{o}wen, C. Reichhardt, G. Volpe, and G. Volpe, Rev. Mod. Phys. {\bf 88}, 045006 (2016).
% Active particles in complex and crowded environments

\bibitem{Ramaswamy2017} S. Ramaswamy, J. Stat. Mech.  054002 (2017).
% Active Matter

\bibitem{Marchetti2017} \'{E}. Fodor, and M. C. Marchetti, Physica A {\bf 504}, 106 (2018). 
% The statistical physics of active matter: from self-catalytic colloids to living cells

\bibitem{Berg2004} {\it E. Coli in Motion}, H. C. Berg,  (Springer Verlag, Heidel-
berg, Germany) (2004).

\bibitem{Cates2012} M. E. Cates, Rep. Prog. Phys. {\bf 75}, 042601 (2012). 
% Diffusive transport without detailed balance: Does microbiology need statistical physics? 

\bibitem{tissue} X. Trepat, M. R. Wasserman, T. E. Angelini, E. Millet, D. A. Weitz, J. P. Butler, and J. J. Fredberg, Nature Physics  {\bf 5}, 426 (2009).
% Physical forces during collective cell migration

\bibitem{Vicsek} T. Vicsek, A. Czir\'{o}k, E. Ben-Jacob, I. Cohen, and O. Shochet, Phys. Rev. Lett. {\bf 75}, 1226 (1995). 
% Novel Type of Phase Transition in a System of Self-Driven Particles

\bibitem{fish} S. Hubbard, P. Babak, S. Th. Sigurdsson, and K. G. Magn\'{u}sson, Ecological Modelling, {\bf 174}, 359 (2004). 
% A model of the formation of fish schools and migrations of fish

\bibitem{gran1} D. L. Blair, T. Neicu, and A. Kudrolli, Phys. Rev. E {\bf 67}, 031303 (2003).
% Vortices in vibrated granular rods

\bibitem{gran2} L. Walsh, C. G. Wagner, S. Schlossberg, C. Olson, A. Baskaran, and N. Menon, Soft Matter {\bf 13}, 8964 (2017). 

\bibitem{cluster2} J. Palacci,  S. Sacanna, A. P. Steinberg, D. J. Pine, and P. M. Chaikin, Science {\bf 339}, 936 (2013). 
% Living crystals of light-activated colloidal surfers 

\bibitem{flocking1} J. Toner, Y. Tu, and S. Ramaswamy, Ann. of Phys. {\bf 318}, 170 (2005).
% Hydrodynamics and phases of flocks

\bibitem{flocking2} N. Kumar, H. Soni, S. Ramaswamy, and A.~K. Sood, Nature Comm. {\bf 5},  4688 (2014). 
% Flocking at a distance in active granular matter


\bibitem{cluster1} Y. Fily, and M. C. Marchetti, \prl {\bf 108}, 235702 (2012).
% Athermal Phase Separation of Self-Propelled Particles with No Alignment

\bibitem{SEB_16} A. B. Slowman, M. R. Evans, R. A. Blythe,
Phys. Rev. Lett. {\bf 116}, 218101 (2016).
%  Jamming and attraction of interacting run-and-tumble random walkers,

\bibitem{SEB_17} A. B. Slowman, M. R. Evans, R. A. Blythe,  J. Phys. A: Math, Theor.  {\bf 50}, 375601 (2017).
% Exact solution of two interacting run-and-tumble random walkers with finite tumble duration

\bibitem{separation1} J. Schwarz-Linek, C. Valeriani, A. Cacciuto, M. E. Cates, D. Marenduzzo, A. N. Morozov, and W. C. K. Poon, Proc. Natl. Acad. Sci. USA {\bf 109}, 4052 (2012). 
% Phase separation and rotor self-assembly in active particle suspensions

\bibitem{separation2} G. S. Redner, M. F. Hagan, and A. Baskaran, Phys. Rev. Lett. {\bf 110}, 055701 (2013).
% Structure and Dynamics of a Phase-Separating Active Colloidal Fluid

\bibitem{separation3} J. Stenhammar, R. Wittkowski, D. Marenduzzo, and M. E. Cates, \prl {\bf 114}, 018301 (2015).
% Activity-Induced Phase Separation and Self-Assembly in Mixtures of Active and Passive Particles

\bibitem{Tailleur2015} A. P. Solon, Y. Fily, A. Baskaran, M. E. Cates, Y. Kafri, M. Kardar, and J. Tailleur, Nature Physics {\bf 11}, 673 (2015).
% Pressure is not a state function for generic active fluids

\bibitem{Potosky2012} A. Pototsky, and H. Stark,  Europhys. Lett. {\bf 98}, 50004 (2012).
% Active Brownian particles in two-dimensional traps

\bibitem{Martens2012} K. Martens, L. Angelani, R. Di Leonardo, and L. Bocquet, Eur. Phys. J. E {\bf 35}, 84 (2012).
% Probability distributions for the run-and-tumble bacterial dynamics: An analogy to the Lorentz model 

\bibitem{ADP_2014} L. Angelani, R. Di Lionardo, and M. Paoluzzi, Euro. J. Phys. E {\bf 37}, 59 (2014).
% First-passage time of run-and-tumble particles

\bibitem{Sevilla2014} F. J. Sevilla, and L. A. G\'{o}mez Nava, Phys. Rev. E {\bf 90}, 022130 (2014).
%Theory of diffusion of active particles that move at constant speed in two dimensions

\bibitem{Solon2015} A.~P. Solon, M.~E. Cates, and J. Tailleur, Eur. Phys. J. Special Topics {\bf 224}, 1231 (2015).
% Active brownian particles and run-and-tumble particles: A comparative study

\bibitem{EG2015} J. Elgeti and  G. Gompper, 
Europhys. Lett. {\bf 109} 58003 (2015).
% Run-and-tumble dynamics of self-propelled particles in confinement,

\bibitem{Angelani15}  L. Angelani, J. Phys. A: Math. Theor. {\bf 48} 495003 (2015).
% Run-and-tumble particles, telegrapher's equation and absorption problems with partially reflecting boundaries.

\bibitem{seifert}
P. Pietzonka, K. Kleinbeck, and U. Seifert, New J. Phys. {\bf 18}, 052001 (2016).
%Extreme fluctuations of active BM 

\bibitem{Takatori} S. C. Takatori, R. De Dier, J. Vermant, and J. F. Brady, Nature Comm. {\bf 7}, 10694 (2016). 
% Acoustic trapping of active matter

\bibitem{Angelani17} L. Angelani, J. Phys. A: Math. Theor. {\bf 50}  325601 (2017).
% Confined run-and-tumble swimmers in one dimension

\bibitem{Malakar2018} K. Malakar, V. Jemseena, A. Kundu, K. Vijay Kumar, S. Sabhapandit, S. N.
Majumdar, S. Redner, A. Dhar, J. Stat. Mech. 043215 (2018).
% Steady state, relaxation and first-passage properties of a run-and-tumble particle in one-dimension

\bibitem{DM_2018} T. Demaerel and C. Maes, Phys. Rev. E {\bf 97}, 032604 (2018).
% Active processes in one dimension

\bibitem{ABP-pre} U. Basu, S. N. Majumdar, A. Rosso, G. Schehr, Phys. Rev. E {\bf 98}, 062121 (2018).
% Active Brownian motion in two dimensions

\bibitem{Franosch} C. Kurzthaler, C. Devailly, J. Arlt, T. Franosch, W. C. K. Poon, V. A. Martinez, and A. T. Brown,
Phys. Rev. Lett. {\bf 121}, 078001 (2018).
% Probing the Spatiotemporal Dynamics of Catalytic Janus Particles with Single-ParticleTracking and Differential Dynamic Microscopy



\bibitem{limmer}
T. GrandPre, and D. T. Limmer, Phys. Rev. E {\bf 98}, 060601(R) (2018). 
%Current fluctuations of interacting active Brownian particles



\bibitem{EM2018} M. R. Evans and S. N. Majumdar, J. Phys. A: Math. Theor. {\bf 51}, 475003 (2018).
% Run and tumble particle under resetting: a renewal approach

\bibitem{Adhar2019} A. Dhar, A. Kundu, S. N. Majumdar, S. Sabhapandit and G. Schehr, 
Phys. Rev. E, {\bf 99}, 032132 (2019).
% Run-and-tumble particle in one-dimensional confining potential: Steady state, relaxation and first passage properties

\bibitem{GM2019} G. Gradenigo and S. N. Majumdar, J. Stat. Mech.  053206 (2019).
% A First-Order Dynamical Transition in the displacement distribution of a Driven Run-and-Tumble Particle

\bibitem{Malakar2019}  K. Malakar, A. Das, A. Kundu, K. Vijay Kumar, A. Dhar, {\it arXiv:1902.04171} 
% Steady State of an Active Brownian Particle in Two-Dimensional Harmonic Trap 

\bibitem{Dauchot2019} O. Dauchot, V. D\'emery, Phys. Rev. Lett. {\bf 122}, 068002 (2019).
% Dynamics of a self-propelled particle in a harmonic trap

\bibitem{LMS2019} P. Le Doussal, S.~N. Majumdar, and G. Schehr,
Phys. Rev. E {\bf 100}, 012113 (2019).
% Non-crossing run-and-tumble particles on a line

\bibitem{SK2019} P. Singh and A. Kundu, J. Stat. Mech. 083205 (2019). 
% Generalised `Arcsine' laws for run-and-tumble particle in one dimension

\bibitem{Sevilla2019} F. J. Sevilla, A. V. Arzola, and E. P. Cital, Phys. Rev. E {\bf 99}, 012145 (2019).
%Stationary superstatistics distributions of trapped run-and-tumble particles

\bibitem{caprini}
L. Caprini, E. Hern{\`a}ndez-Garc"a, C. L{\`o}pez, and U. M. B. Marconi,  preprint arXiv:1906.03016.
%A comparative study between two models of active cluster-crystals

\bibitem{Mumford} D. Mumford,  {\it Elastica and Computer Vision}, in  Algebraic Geometry and its Applications, edited by  C. L. Bajaj, Springer, New York (1994)

\bibitem{satya_review} S. N. Majumdar, Curr. Sci. {\bf 77}, 370 (1999).
% Persistence in Nonequilibrium Systems

\bibitem{Mathieu} M. Abramowitz and I. A. Stegun (Eds.). {\it Handbook of Mathematical Functions with 
Formulas, Graphs, and Mathematical Tables}, 9th printing. New York: Dover, p. 928, 1972. 



\bibitem{LS99}
J. L. Lebowitz, and H. Spohn, J. Stat. Phys. {\bf 95}, 333 (1999).
%A GallavottiÐCohen-type symmetry in the large deviation functional for stochastic dynamics

\bibitem{IS1} A. K. Hartmann, Phys. Rev. E {\bf 65}, 056102 (2002).
% Sampling rare events: Statistics of local sequence alignments

\bibitem{IS2} A. K. Hartmann, Eur. Phys. J. B {\bf 84}, 627 (2011).
% Large-deviation properties of largest component for random graphs

\bibitem{ISRMT1} C. Nadal, S. N. Majumdar and M. Vergassola, J. Stat. Phys., {\bf 142}, 403 (2011).
% Statistical Distribution of Quantum Entanglement for a Random Bipartite State

\bibitem{ISRMT2} S. N. Majumdar, C. Nadal, A. Scardicchio, and P. Vivo, Phys. Rev. E, {\bf 83}, 041105 (2011).
% How many eigenvalues of a Gaussian random matrix are positive?

\bibitem{ISCH1} G. Claussen, A. K. Hartmann, and S.~N. Majumdar, Phys. Rev. E, {\bf 91}, 052104 (2015).
% Convex hulls of random walks: Large-deviation properties

\bibitem{ISCH2} T. Dewenter, G. Claussen, A. K. Hartmann, and S.~N. Majumdar, Phys. Rev. E, {\bf 94}, 052120 (2016).
% Convex hulls of multiple random walks: a large-deviation study

\bibitem{ISCH3} H. Schawe, A. K. Hartmann, S. N. Majumdar, Phys. Rev. E  {\bf 96}, 062101 (2017).
% Convex Hulls of Random Walks in Higher Dimensions: A Large Deviation Study

\bibitem{ISCH4} H. Schawe, A. K. Hartmann, S. N. Majumdar, Phys. Rev. E  {\bf 97}, 062159 (2018).
% Large Deviations of Convex Hulls of Self-Avoiding Random Walks

\bibitem{ISCH5} H. Schawe and A. K. Hartmann, Eur. Phys. J. B {\bf 92}, 73 (2019).
% Large-deviation properties of the largest biconnected component for random graphs

\bibitem{ISKPZ2} A. K. Hartmann, P. Le Doussal, S. N. Majumdar, A. Rosso, G. Schehr, Europhys. Lett. {\bf 121}, 67004 (2018).
% High-precision simulation of the height distribution for the KPZ equation

\bibitem{ISLIS} J. Borjes, H. Schawe, and A. K. Hartmann,  Phys. Rev. E {\bf 99}, 042104 (2019).
%Large deviations of the length of the longest increasing subsequence of random permutations and random walks

\bibitem{Gredat} D. Gredat, I. Dornic, and J.~M. Luck,  J. Phys. A: Math. Theor. {\bf 44},  175003 (2011).
% On an imaginary exponential functional of Brownian motion

\bibitem{footnote1} Note that, in polar coordinates, the normalization of
the total probability translates to 
$\int_0^\infty r\, P_{\textrm{rad}}(r,t) \, dr = 1 $.

\bibitem{footnote2} To be convinced one can simply plot $g_z(h)$ for
different values of $z$.



\end{thebibliography}
\end{document}